\documentclass[journal]{IEEEtran}
\usepackage{cite}
\usepackage{graphicx,amssymb,lineno}
\usepackage{amsmath,amsfonts,amssymb}

\usepackage{algorithm}
\usepackage{algorithmic}
\usepackage[usenames]{color}
\usepackage{float}
\usepackage{mathtools}

\usepackage{graphicx,graphics,color,epsfig,subfigure,graphpap,rotate}
\usepackage{times, verbatim, subfigure, epsfig, graphicx, latexsym, amsmath}
\usepackage{url}
\usepackage{subfigure}
\usepackage{CJK}
\begin{document}

\title{Energy-Efficient Power Control of Train-ground mmWave Communication for High Speed Trains}

\author{Lei~Wang,
        Bo~Ai,~\IEEEmembership{Senior Member,~IEEE},
        Yong~Niu,
        Xia~Chen,
        and Pan~Hui,~\IEEEmembership{Fellow,~IEEE}
\thanks{L.~Wang, B.~Ai, Y.~Niu, and X.~Chen are with the State Key Laboratory of Rail Traffic Control and Safety, Beijing Jiaotong University, Beijing 100044, China (E-mails: 15801531622@163.com; niuy11@163.com; boai@bjtu.edu.cn; xchen@bjtu.edu.cn).}

\thanks{P.~Hui is with the Department of Computer Science and Engineering, The Hong Kong University of Science and Technology, Hong Kong, and Department of Computer Science and Engineering, University of Helsinki, Finland.}

\thanks{Copyright(c) 2015 IEEE. Personal use of this material is permitted. However, permission to use this material for any other purposes must be obtained from the IEEE by sending a request to pubs-permissions@ieee.org.}

\thanks{This study was supported by National Key R\&D Program of China under Grant 2016YFE0200900; and by the National Natural Science Foundation of China Grants 61725101, 61801016, and U1834210; and by the China Postdoctoral Science Foundation under Grant 2017M610040 and 2018T110041; and by the Beijing Natural Fund under Grant L172020; and by Major projects of Beijing Municipal Science and Technology Commission under Grant No. Z181100003218010; and by project 16214817 from the Research Grants Council of Hong Kong and the 5GEAR project from the Academy of Finland. (\emph{Corresponding authors: B. Ai, Y. Niu.})}
}
\maketitle

\begin{abstract}
High speed train system has proven to be a very flexible and attractive system that can be developed under various circumstances and in different contexts and cultures. As a result, high speed trains are widely deployed around the world. Providing more reliable and higher data rate communication services for high speed trains has become one of the most urgent challenges. With vast amounts of spectrum available, the millimeter wave (mmWave) system is able to provide transmission rates of several gigabits per second for high speed trains. At the same time, mmWave communication also suffers from high attenuation, thus higher energy efficiency should be considered. This paper proposes an energy efficient power control scheme of train-ground mmWave communication for high speed trains. Considering a beam switching method for efficient beam alignment, we first establish position prediction model, the realistic direction antenna model and receive power model. And then we allocate the transmission power rationally through the power minimization algorithm while ensuring the total amount of transmission data. Based on this, this paper also develops a hybrid optimization scheme and finds the limit of total energy consumption when the number of segments goes to infinity. Through simulation with various system parameters and taking velocity estimation error into account, we demonstrate the superior performance of our schemes.
\end{abstract}

\section{Introduction}\label{S1}
High speed railway is now a mature system of transport. And high speed train (HST) systems are considered as a green transportation system that can provide fast and convenient services. HSTs have been widely deployed around the world and the network is still rapidly expanding. It is expected that the total length will reach up to 54,550 km by 2025 \cite{Qualcomm}. Providing quality broadband services for passengers is becoming especially important. With the increasing demand of bandwidth, the current 4G wireless technology cannot meet the needs of passengers to access the Internet using mobile terminals in a high-speed mobile environment. The contradiction between the increasing demand for transmission data of train-ground communication and the existing low-rate communication makes it urgent to solve the problem of wireless broadband communication for HSTs.

As a technology adopted by the fifth-generation (5G) wireless communication,  mmWave communication is able to support multi-gigabit wireless
services like wireless gaming, wireless Gigabit Ethernet, real-time compressed and uncompressed high-definition television (HDTV) streaming media, and high-speed data transfer between devices such as cameras, tablets and personal computers \cite{2gai2}. Therefore, it can solve the problem of train-to-ground wireless broadband transmission effectively and provide stable and reliable communication services for passengers, which is difficult for existing operators in a high-speed mobile environment. At present, the technology has been applied in some HST communication systems worldwide and have achieved good performance, like the Japanese Shinkansen, the TVE test line of maglev train in Germany and the maglev train in Shanghai, China \cite{Ruisi}. Due to its wide bandwidth, narrow beam and rich spectrum resources, applying mmWave to train-ground communication for HSTs not only conforms to the current development trend of wireless communication networks, but also satisfies the needs of passengers for Internet access. It has become an effective way to improve the quality of HST services.

For HST communication systems, although many new technologies have been proposed, there are still many important issues need to be investigated. Due to high carrier frequency, mmWave communications suffer from higher propagation loss than communication systems in lower frequency bands \cite{gai7}. For instance, the free space path loss at 60 GHz band is 28 decibels (dB) higher than that at 2.4 GHz \cite{2gai1}. With the maturity of directional antenna technology \cite{spaceair}, mmWave communication has attracts considerable attention. To combat high channel attenuation, analog beamforming techniques are exploited to synthesize directional antennas at both the transmitter and receiver to achieve high antenna gain \cite{gai8,gai9}. Consequently, the omnidirectional carrier sensing is disabled, and this is the deafness problem \cite{laoshi1}.

Energy problem is another important problem that needs to be addressed. Over the past years, the work on 5G networks has achieved remarkable results \cite{01101}, \cite{01102}. 3GPP has recently ratified non-standalone 5G New Radio (NR) technology to augment further LTE \cite{01103}. In the 5G era, with a large number of base stations (BSs) deployed, the cost-efficient, flexible, and green power control solution becomes one of the most urgent and critical challenges \cite{niu2017energy}. The energy efficiency of train-ground communication for HSTs has gradually become a key issue in green wireless communication. First, the HST system is divided into five subsystems and energy system is one of them, enhancing energy efficiency can help improve the performance of the HST system. And high energy efficiency is one of goals of future HST developments, with higher energy efficiency, the HST system will be more sophisticated \cite{energy1}. Second, as a mass transport system, the HST system has many fixed costs and huge investment required, it needs specific ground infrastructure which is costly to implement and maintain. In order to fulfil its potential in meeting future transport needs, the rail industry will need to radically progress in terms of costs. Reduce energy consumption is beneficial to reduce operations costs to some extent \cite{energy2}. Finally, compared to the road and air industries, the rail industry is small, improvement of energy efficiency can help the rail industry keep pace with road and air. Up until now, rail has been considered the safest and most environmentally friendly transport mode. However, driverless electric cars will completely transform how we view road transport. Similar leaps forward are anticipated in the air industry, towards the production of much quieter and more fuel efficient airplanes. The rail industry cannot sit back and simply watch all of these breakthroughs happen. Even where rail transport is head and shoulders ahead, improvements and innovations are of the utmost necessity \cite{why}. Therefore, there is strong motivation and interest to investigate an energy-efficient solution for HST wireless communication systems.

This paper considers a HST network of mmWave access points similar to the architecture proposed in \cite{IEEE802153c}. We focus on a mmWave system that employs directional beamforming at the transmitter (TX) and receiver (RX), using adaptive arrays. The IEEE 802.11 ad/ay standard has adopted the way of beamforming to increase antenna gain \cite{FiWi}. A beam switching method is considered which uses train position information similar to \cite{mmWavesub6}, and the information can be obtained from the train control system (TCS). Modern railway systems have TCSs, which can track the position and velocity of each train. Assuming that the BSs and the transceivers on the train (mobile relay) have only one RF chain, the BS uses mmWave directional transmission to the mobile relay on the train, and beamforming is accomplished via a digitally controlled phased array. With the estimated position information of the train, the time to switch the beam direction can be calculated. According to the result of \cite{IEEE802153c}, this switching can be very fast ($<$50ns).

In this paper, we develop energy-efficient schemes of train-ground mmWave communication for HSTs, where the transmission power is allocated rationally according to the distance between the train and the BS. Based on the power minimization algorithm, our schemes achieve lower energy consumption and establish an energy-efficient communication mechanism. To the best of our knowledge, we are the first to analyze the energy consumption problem of train-ground mmWave communication by dividing the inter-BS distance into small segments. Besides, we also find the limit value when the number of segments goes to infinity. The scheme is based on a beam switching method which can achieve efficient beam alignment. The contributions of this paper are summarized as follows.

\begin{itemize}
\item We focus on the energy efficiency power control problem of train-ground mmWave communication and formulate the problem into a nonlinear optimization problem, the optimization model is established based on position prediction model, the realistic direction antenna model and receive power model. And the position prediction model is established based on the beam switching scheme.

\item We propose an energy-efficient power control scheme where the coverage of a base station is divided into small segments and the transmission power is allocated properly through power minimization algorithm. The total amount of transmission data is considered in the problem. We also develop a hybrid optimization scheme that only allocates the transmission power of the second half of the considered part. Furthermore, we find the limit of total energy consumption when the number of segments approaches infinity.

\item Through simulation with various system parameters and taking velocity estimation error into account, we demonstrate that our schemes can achieve lower energy consumption and higher energy efficiency.
\end{itemize}

The rest of this paper is organized as follows. Section~\ref{S2} introduces the related work on the mmWave communication for HSTs. Section~\ref{S3} presents the system model and formulates the energy efficiency optimization problem of train-ground mmWave communication into a nonlinear minimization problem. Section~\ref{S4} proposes our energy-efficient power control schemes. In section~\ref{S5}, the performance of the scheme is analyzed when the number of segments tends to infinity. Section~\ref{S6} gives the evaluation of our schemes in terms of energy consumption and energy efficiency with velocity estimation error and various system parameters. Section~\ref{S7} concludes this paper.

\section{Related Work}\label{S2}

There have been some schemes proposed on the mmWave communication system for HSTs \cite{singh2011interference,jiajia1,YongNiu,2gai4}. Va \emph{et al}. \cite{YongNiu} considered a beam switching method which uses the position information of the train from TCS to determine the time of beam switching, velocity estimation error is also taken into account to get the optimal beamwidth. Kim \emph{et al}. \cite{singh2011interference} proposed a distributed antenna system-based (DAS) mmWave communication system for HSTs based on the hierarchical two-hop network, antenna modules are distributed by geographic position and the communication interruption time during handover was minimized to provide more reliable links between TX and RX. In \cite{jiajia1}, a disaster radar detection approach for the safety of trains was designed. According to the safety level, the area around the railway is divided into three parts and different detection methods are used for different sub-areas. \cite{2gai4} introduced orthogonal frequency-division multiplexing (OFDM) and single carrier (SC) scheme to support mmWave communication and made a suggestion that train-trackside mmWave systems should deploy both OFDM and SC, and then presented train-trackside network architecture adopted MIMO technologies.

The upcoming 5G mobile communication system is expected to support high mobility up to 500 km/h, which is envisioned in particular for HSTs. MmWave spectrum is considered as a key enabler for offering the ``best experience'' to highly mobile users \cite{jiajia2,jiajia3,ma2015interference,ramezani2017joint,2gai3}. In order to solve the problem of applying traditional MIMO to HSR scenario, Cui \emph{et al}. \cite{jiajia2} proposed a hybrid beamforming scheme using spatial modulation (SM) which can achieve multiple antenna gain, and the scheme has been designed both in analog domain and in digital domain. \cite{jiajia3} considered the mobile hotspot network based on a hierarchical relay network structure, the design of baseband modem and RF front end are given which can satisfy high data transfer rates for HSTs with the velocity up to 500km/h, and the core of system structure is mitigating severe Doppler effects by allocating downlink and uplink reference signal. \cite{ma2015interference} discussed the main challenges of 5G mmWave HST communication and proposed a viable paradigm by definiting the 5G mmWave HST scenario and selecting proper objects and materials. Through reconstruction of three models, the paradigm is verified. Talvitie \emph{et al}. \cite{ramezani2017joint} studied HST positioning problem in a 5G NR network by using specific NR synchronization signals. The positioning method utilizes measurement of time of arrival and angle of departure, and the position can be tracked by an Extended Kalman Filter. \cite{2gai3} overcame the challenge of sensitivity of mmWave links by utilizing multiple antennas diversity and higher bandwidth. A risk-sensitive reinforcement learning framework was formulated where each cell optimizes its transmission while considering signal fluctuations.

Recently, there are also some works on mmWave beam alignment and power allocation \cite{1,21,3,4,5}. To realize fast and accurate estimation of mmWave channel, Xiao \emph{et al}. \cite{1} proposed a multipath decomposition and recovery approach by exploiting the spacial sparsity. Particularly, a codebook is designed for the approach to make it applicable for both analog and hybrid beamforming/combining devices with strict constantmodulus(CM) constraint. Considering the non-orthogonal multiple access in mmWave communications, \cite{21} studied power and beam gain allocation problem and beamforming problem under the CM constraint to maximize the sum rate of a 2-user mm-wave-NOMA system. Zhou \emph{et al}. \cite{3} investigated the problem of beam misalignment at both the base station and the users and developed a performance analysis framework for mmWave-NOMA networks with spatially random users. Liu \emph{et al}. \cite{4} formulated the discrete power control and non-unified transmission duration allocation problem for self-backhauling mmWave cellular networks as an optimization problem and corresponding algorithms have been designed to solve it. Cui \emph{et al}. \cite{5} proposed a branch and bound (BB) based power allocation algorithm and a match theory based algorithm to maximize the sum rate for the mmWave NOMA system.

To the best of our knowledge, most of these works do not focus on the energy consumption reduction problem of train-ground mmWave communication for HSTs, and energy efficiency problem is not considered. In this paper, we investigate the problem and allocate the transmission power properly to achieve good performance.

\section{System Model and Problem Formulation}\label{S3}

This section describes problem formulations and the models needed for energy efficiency optimization.

\subsection{Position Prediction Model}\label{S3-1}

\begin{figure}[htp]
\begin{center}
\includegraphics*[width=0.9\columnwidth,height=1.4in]{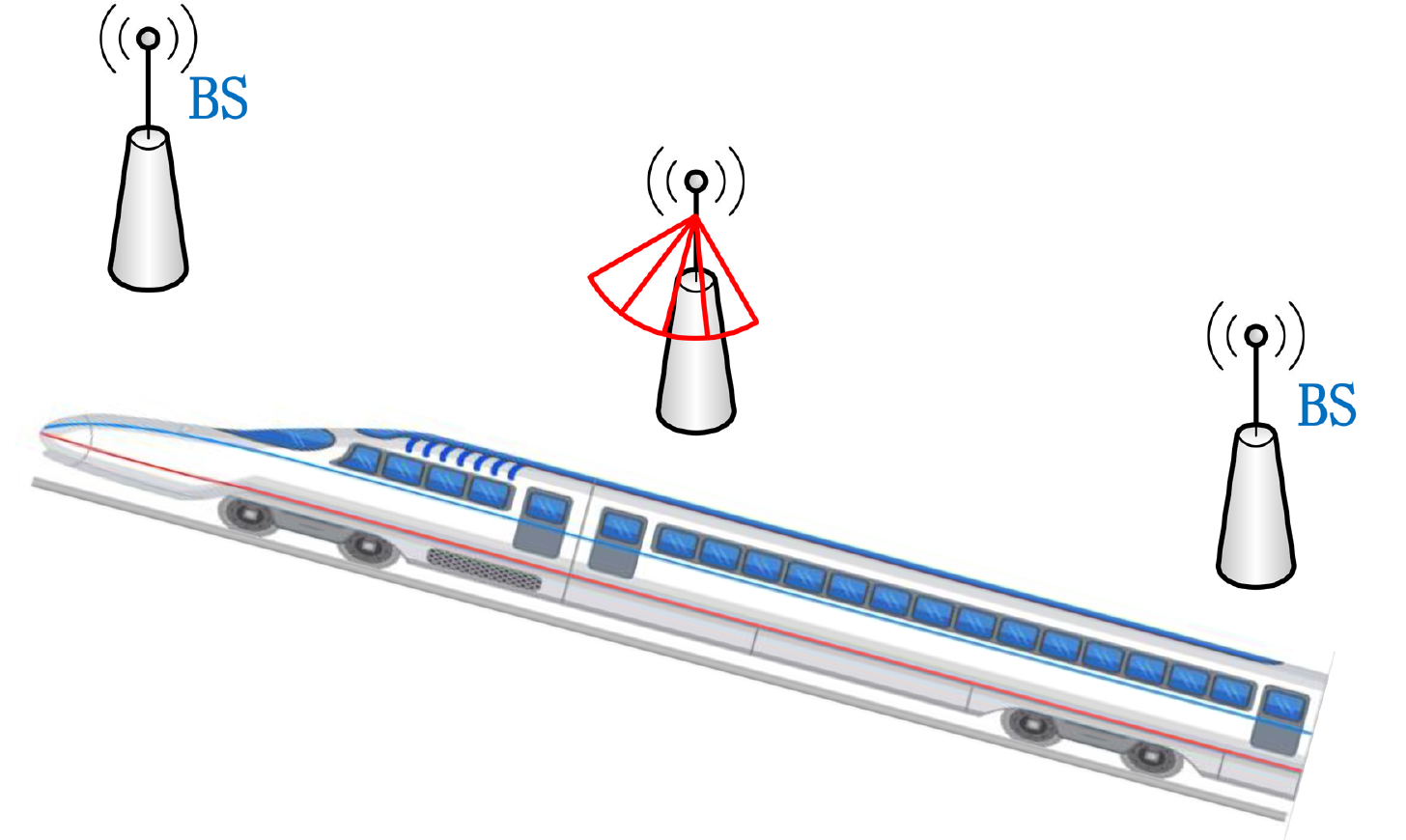}
\end{center}
\caption{A mmWave network for HST.}
\label{fig1}
\end{figure}

\begin{figure}[htp]
\begin{center}
\includegraphics*[width=0.8\columnwidth,height=1.2in]{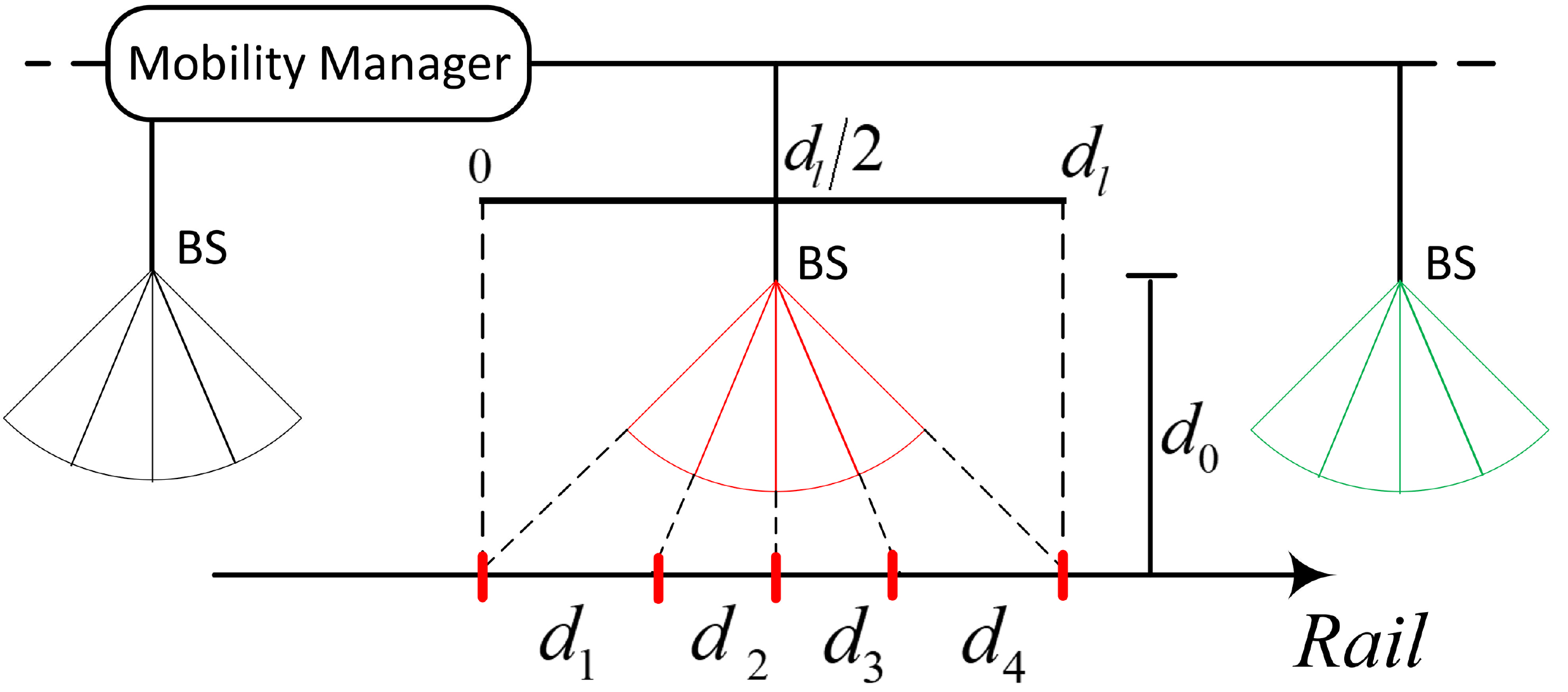}
\end{center}
\caption{Bird view of the mmWave network.}
\label{fig2}
\end{figure}

Consider a network model as shown in Fig. \ref{fig2}, where the distance from a BS to the rail is denoted by $d_0$ and the inter-BS distance by $d_l$. We assume that the total number of segments is even and from symmetry, we only need to consider the half of the model, \emph{i.e.}, the coverage of a BS is divided into $2N$ beams ($N=2$ in Fig. \ref{fig2}), and the coverage of the $i^{th}$ beam is $d_i$, $d_i$ can be computed from the geometry.

We assume that only periodic feedback from TCS can provide an accurate current position and a velocity estimate, the communication system uses this information to predict the position of the train until the next update is available. The predicted positions are used to determine the time of beam switching.

Assume that the speed of the train is constant, notice that trains with large mass cannot accelerate or decelerate rapidly, so this assumption is reasonable. Then the position can be modeled as
\begin{equation}
x(t)=v(t-t_0)+x_0, \label{eq1}
\end{equation}
where $x_0$ is the feedback location, and $t_0$ is the time of the feedback. Without loss of generality, we set $x_0=0$ and $t_0=0$.

\subsection{The Realistic Direction Antenna Model}\label{S3-2}

In this paper, we adopt the widely used realistic directional antenna model, which is a main lobe of Gaussian form in linear scale and constant level of sidelobes \cite{2}. It is the reference antenna model with sidelobe for IEEE 802.15.3c. And we assume that the Doppler effect of the channel has been eliminated by modulation methods \cite{OTFS}. The gain of a directional antenna in units of decibels $(dB)$, which is denoted by $G(\theta)$, can be expressed as
\begin{equation}
G(\theta)=\left\{
\begin{array}{rcl}
G_0-3.01\cdot{(\frac{2\theta}{\theta_{-3dB}})}^{2},&{0^\circ\leq\theta\leq{\theta_{ml}}/2}\\
G_{sl},&{{\theta_{ml}}/2\leq\theta\leq180^\circ}
\end{array}
\right.,\label{eq2}
\end{equation}
where $\theta$ denotes an arbitrary angle within the range $\left[0,180^\circ\right]$, $\theta_{-3dB}$ denotes the angle of the half-power beamwidth, and $\theta_{ml}$ denotes the main lobe width in units of degrees. The relationship between $\theta_{ml}$ and $\theta_{-3dB}$ is $\theta_{ml}=2.6\cdot\theta_{-3dB}$.
$G_0$ is the maximum antenna gain, and this can be obtained by
\begin{equation}
G_0=10\log{\left(\frac{1.6162}{\sin{\left(\theta_{-3dB}/2\right)}}\right)^2},\label{eq3}
\end{equation}
The sidelobe gain $G_{sl}$ can be expressed as
\begin{equation}
G_{sl}=-0.4111\cdot{\ln(\theta_{-3dB})-10.579}.\label{eq4}
\end{equation}

\subsection{Receive Power Model}\label{S3-3}

The RX power is modeled as \cite{mmWavesub6}
\begin{equation}
P_{rx}^{dBm}=P_{tx}^{dBm}+G_{tx}+G_{rx}-W+10n\log_{10}{\frac{\lambda}{4\pi{d}}},\label{eq5}
\end{equation}
where, $P_{rx}^{dBm}$ and $P_{tx}^{dBm}$ are the RX and TX powers, $G_{rx}$ and $G_{tx}$ are the RX and TX antenna gains, $W$ is the shadowing margin, $\lambda$ is the carrier wavelength, and $d$ is the distance from the BS to the mobile relay on the train. We can approximate $G_{rx}$ and $G_{tx}$ with $G_0$, and compute the $d$ as a function of time, then the RX power is modeled as
\begin{equation}
P_{rx}^{dBm}(t)=P_{tx}^{dBm}+2G_{0}-W+10n\log_{10}{\frac{\lambda}{4\pi{d(t)}}}.\label{eq6}
\end{equation}

Denoting $B$ the system bandwidth, and NF the noise figure of the receiver chain, we model thermal noise as
\begin{equation}
P_{noise}^{dBm}=-174+10\log_{10}{B}+\mathrm{NF}.\label{eq7}
\end{equation}

The received SNR is determined by not only the received signal power but also the noise power and can be expressed as \cite{22,24}
\begin{equation}
SNR(t)=\frac{P_{rx}(t)}{P_{noise}}.\label{eq8}
\end{equation}

\subsection{Problem Formulation}\label{S3-4}

As we know, the transmission power allocation is a key mechanism to energy consumption, if the transmission power can be allocated properly, the energy consumption will be reduced greatly, therefore, the transmission power should be optimized to achieve high energy efficiency. Here, we formulate the transmission power optimization problem to minimize energy consumption.

The energy efficiency optimization problem of train-ground mmWave communication can be formulated as
\begin{equation}
\max\ EE,\label{eq9}
\end{equation}
where $EE$ denotes the energy efficiency and can be expressed as
\begin{equation}
EE=\frac{D}{E},\label{eq10}
\end{equation}
D denotes the total amount of transmission data and the total energy consumption is $E$.

When maximizing the energy efficiency, we only need to minimize the energy consumption and ensure that the amount of data is greater than a certain value simultaneously, then the problem of (\ref{eq10}) can be transformed as follows:
\begin{equation}
\left\{
\begin{array}{lr}
\min\quad E & \\
s.t.\quad D\ge{D_{fixed}} &
\end{array}
\right.,\label{eq11}
\end{equation}

Assume that the transmission power at each segment is a constant $P_i$, then the total energy consumption can be expressed as $E=\sum\limits_{i} P_it$.

Notice that $t=\displaystyle\frac{d_i}{v}$, the problem of (\ref{eq11}) can be expressed as
\begin{equation}
\left\{
\begin{array}{lr}
\min\quad \sum\limits_{i=1}^N \displaystyle\frac{d_i}{v}P_i & \\
s.t.\quad D\ge{D_{fixed}} &
\end{array}
\right..\label{eq12}
\end{equation}

This is a nonlinear minimization problem where objective function is a linear function about $P_i$ and both $d_i$ and $v$ can be seen as a constant, and the constraint indicates that $P_i\ge0,i\in[1,N]$.

To reduce the energy consumption, we should ensure the total amount of transmission data. Now, we analyze the system constraint of this optimization problem. First, according to the Shannon capacity formula, the achievable data rate is determined by the received SNR \cite{25,23}, and the instantaneous rate is formulated as
\begin{equation}
R(t)=\log_{2}{(1+SNR(t))},\label{eq13}
\end{equation}

Based on the linear position prediction model, the system will switch to the $i^{th}$ beam at time $\sum_{j=1}^{i-1} d_j/v$ and keeps this beam
until time $\sum_{j=1}^{i} d_j/v$, then the total amount of transmission data of the $i^{th}$ beam is

\begin{equation}
D_i=\int_{\sum\limits_{j=1}^{i-1} d_j/v}^{\sum\limits_{j=1}^i d_j/v} \log_{2}{(1+SNR(t))}\, dt,\label{eq14}
\end{equation}

\begin{figure}[htp]
\begin{center}
\includegraphics*[width=0.7\columnwidth,height=1.4in]{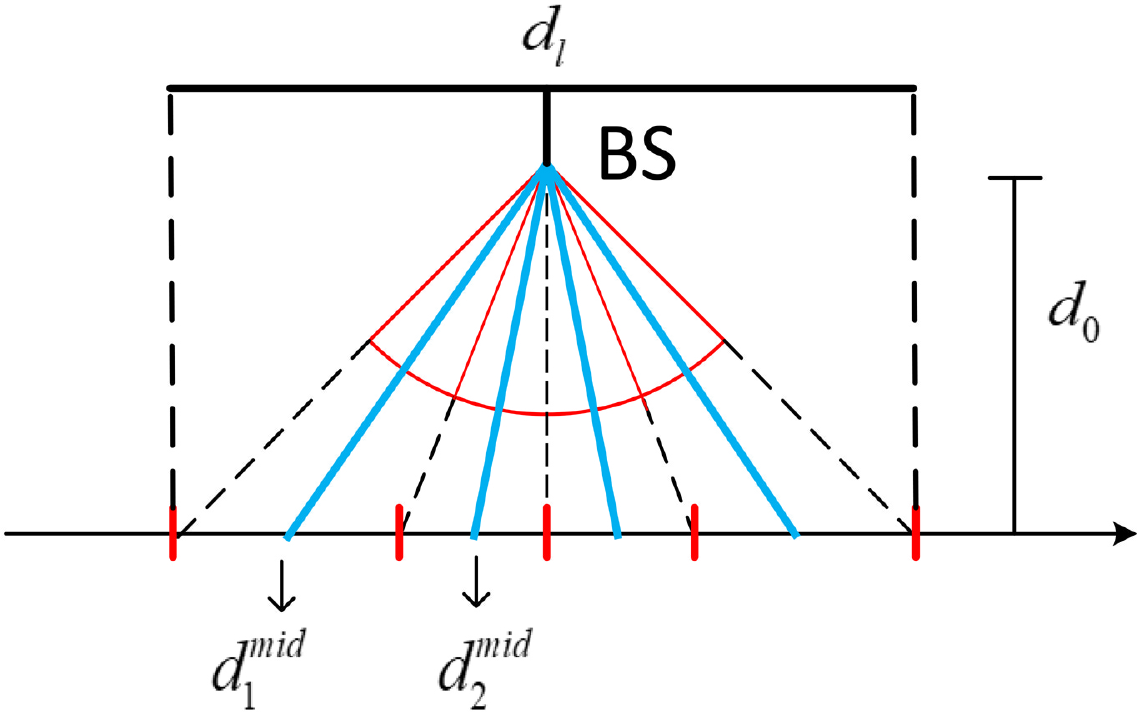}
\end{center}
\caption{Midpoint approximation of the network.}
\label{fig3}
\end{figure}

Approximate the SNR of the segment with it at the midpoint of this segment, as shown in Fig. \ref{fig3}, then $D_i$ can be estimated as
\begin{equation}
D_i\approx\log_{2}{(1+SNR_i^{mid})}\cdot\frac{d_i^{mid}}{v},\label{eq15}
\end{equation}
where, $SNR_i^{mid}$ is the SNR at the midpoint of the $i^{th}$ segment and $d_i^{mid}$ is the distance between BS and the midpoint of each segment of the rail.

According to (\ref{eq8}),
\begin{equation}
SNR_i^{mid}=\frac{P_i+2G_{0}-W+10n\log_{10}{\frac{\lambda}{4\pi{d_i^{mid}}}}}{-174+10\log_{10}{B}+\mathrm{NF}},\label{eq16}
\end{equation}

From the geometry,
\begin{equation}
d_i^{mid}=\left[(\frac{d_l}{2}-\sum_{j=1}^{i-1} d_j-\frac{d_i}{2})^2+d_0^2\right]^{\frac{1}{2}},\label{eq17}
\end{equation}

From the approximate formula, we can get that the total amount of transmission data is $D=\sum\limits_{i=1}^{2N} d_i$.

For problem of (\ref{eq12}),
\begin{equation}
\begin{aligned}
D&=\sum_{i=1}^ND_i\\
&\approx\sum_{i=1}^N \log_{2}{(1+SNR_i^{mid})}\cdot\frac{d_i^{mid}}{v},\label{eq18}
\end{aligned}
\end{equation}

From the geometry, the inter-BS distance $d_i$ can be calculated as
\begin{equation}
d_i=d_0[\tan(N+1-i)\theta-\tan(N-i)\theta],i\in[1,N],\label{eq19}
\end{equation}

$\theta$ is determined by $N$ and can be obtained by
\begin{equation}
\theta=\frac{\arctan\left(\displaystyle\frac{d_l}{2d_0}\right)}{N}.\label{eq20}
\end{equation}

Second, when transmission power is a constant, that is, transmission power of each segment is the same, we denote it as $P$, and then $D_{fixed}$ can be calculated as
\begin{equation}
D_{fixed}=\int_{0}^{\frac{d_l}{2v}} \log_{2}{(1+SNR(t))}\, dt,\label{eq21}
\end{equation}
where
\begin{equation}
SNR(t)=\frac{P+2G_{0}-W+10n\log_{10}{\frac{\lambda}{4\pi{d(t)}}}}{-174+10\log_{10}{B}+\mathrm{NF}}, \label{eq22}
\end{equation}
\begin{equation}
d(t)={\left[d_0^2+{\left(\frac{d_l}{2}-vt\right)}^2\right]}^{\frac{1}{2}}.\label{eq23}
\end{equation}

From above, we formulate the energy efficiency optimization problem and analyze its constraint, in the next section, the power minimization algorithm is proposed to solve problem (\ref{eq12}) with low complexity.

\section{Power Control Algorithm}\label{S4}

Here, we propose a power minimization algorithm for the formulated problem which can achieve rational power allocation. The main idea is using Lagrangian multiplier method and based on the algorithm, we adjust the transmission power and get the final result.

To be simplified, the power minimization problem in this paper can be expressed as
\begin{equation}
\left\{
\begin{array}{lr}
\min\quad \sum\limits_{i=1}^N a_iP_i & \\
s.t.\quad \sum\limits_{i=1}^N \log_{2}{(1+c_i+gP_i)}\ge{D_{fixed}} &
\end{array}
\right.,\label{eq25}
\end{equation}

where
\begin{equation}
a_i=\frac{d_i}{v},\label{eq26}
\end{equation}
\begin{equation}
g=\frac{1}{-174+10\log_{10}{B}+\mathrm{NF}},\label{eq27}
\end{equation}
\begin{equation}
c_i=\frac{2G_{0}-W+10n\log_{10}{\frac{\lambda}{4\pi{d_i^{mid}}}}}{-174+10\log_{10}{B}+\mathrm{NF}}.\label{eq28}
\end{equation}

To solve this problem, we use Lagrangian multiplier method and construct a Lagrangian function as follows:
\begin{equation}
L=\sum_{i=1}^N a_iP_i-\lambda\left[\sum_{i=1}^N \log_{2}{(1+c_i+gP_i)}-D_{fixed}\right],\label{eq29}
\end{equation}

Finding partial derivative of variable $P_i$ and then we get the optimal solution. Let $\displaystyle\frac{\partial L}{\partial P_i}=0$ and $\displaystyle\frac{\mathrm{d}L}{\mathrm{d}\lambda}=0$, then
\begin{equation}
P_i=\frac{2^{\frac{D_{fixed}}{a_i\cdot{N}}}-1-c_i}{g}.\label{eq30}
\end{equation}

From (\ref{eq30}), we can obtain all transmission power allocation clearly.

\section{Performance Analysis}\label{S5}

From the analysis above, we can see that $N$ determines $\theta$, and $d_i$ is related to $N$, besides, the value of $N$ is larger, $d_i^{mid}$ and $SNR_i^{mid}$ are more accurate. Since the number of segments $N$ has a big impact on our schemes, here we analyze the energy consumption when $N$ tends to infinity.

To be simplified, let
\begin{equation}
\varphi_{i}(\theta)=\tan(N+1-i)\theta-\tan(N-i)\theta,\label{eq31}
\end{equation}
\begin{equation}
Q=\frac{v\cdot D_{fixed}}{d_0},\label{eq32}
\end{equation}
\begin{equation}
M=2G_0-W,\label{eq33}
\end{equation}

Then the energy consumption can be calculated as
\begin{equation}
\begin{aligned}
\begin{split}
E&=\sum_{i=1}^N \frac{d_i}{v}P_i\\
&=\frac{d_0}{v}\sum_{i=1}^N \varphi_{i}(\theta)\frac{2^{\frac{D_{fixed}}{a_i\cdot{N}}}-1-c_i}{g}\\
&=\frac{d_0}{vg}\sum_{i=1}^N\varphi_{i}(\theta)\cdot 2^{\frac{Q}{\varphi_{i}(\theta)\cdot N}}\\
&-\frac{d_0}{v}\sum_{i=1}^N\varphi_{i}(\theta)\left(\frac{1}{g}+M+10n\log_{10} \frac{\lambda}{4\pi}\right)\\
&+\frac{d_0}{v}\sum_{i=1}^N\varphi_{i}(\theta)\cdot 5n \log_{10} \left((\frac{d_l}{2}-\sum_{j=1}^{i-1} d_j-\frac{d_i}{2})^2+d_0^2\right).\label{eq34}
\end{split}
\end{aligned}
\end{equation}

The formula contains three parts and we take the limits separately to obtain the final result.

For the first part, we can obtain that
\begin{equation}
\lim_{N \to \infty}\sum_{i=1}^N\varphi_{i}(\theta)\cdot 2^{\frac{Q}{\varphi_{i}(\theta)\cdot N}}=H+K(2^{\frac{Q}{K}}-1),\label{eq35}
\end{equation}
where
\begin{equation}
H=\frac{d_l}{2d_0}=\tan N\theta=\sum_{i=1}^N \varphi_{i}(\theta),\label{eq36}
\end{equation}
\begin{equation}
K=\arctan H=N\theta.\label{eq37}
\end{equation}

For the second part, we can obtain that
\begin{equation}
\begin{aligned}
\begin{split}
&\lim_{N \to \infty}\sum_{i=1}^N\varphi_{i}(\theta)\left(\frac{1}{g}+M+10n\log_{10} \frac{\lambda}{4\pi}\right)\\
&=H\left(\frac{1}{g}+M+10n\log_{10} \frac{\lambda}{4\pi}\right).\label{eq38}
\end{split}
\end{aligned}
\end{equation}

For the last part, we can obtain that
\begin{equation}
\begin{aligned}
\begin{split}
&\lim_{N \to \infty}\sum_{i=1}^N\varphi_{i}(\theta)\cdot 5n \log_{10} \left((\frac{d_l}{2}-\sum_{j=1}^{i-1} d_j-\frac{d_i}{2})^2+d_0^2\right)\\
&=H\cdot 10n \log_{10} d_0.\label{eq39}
\end{split}
\end{aligned}
\end{equation}

Then we can get the result that
\begin{equation}
\begin{aligned}
\begin{split}
&\lim_{N \to \infty}E\\
&=\lim_{N \to \infty}\sum_{i=1}^N \frac{d_i}{v}P_i\\
&=\frac{d_0}{v}\left[H\left(10n\frac{4\pi d_0}{\lambda}-M\right)+\frac{K}{g}\left(2^{\frac{Q}{K}}-1\right)\right].\label{eq40}
\end{split}
\end{aligned}
\end{equation}

From (\ref{eq40}), we can obtain the limit value of energy consumption and it should be noted that with the increase of $N$, the value will become smaller and tends to be fixed finally.

\section{Performance Evaluation}\label{S6}

In this section, we evaluate the performance of our energy efficiency schemes under various system parameters and we also analyze the situation which takes velocity estimation error into account. Specifically, we make a comparison of four schemes to show the optimization effect.

\begin{table}[htp]
\begin{center}
\caption{SIMULATION PARAMETERS}
\begin{tabular}{ccc}
\hline
Parameter & Symbol  & Value \\
\hline
Distance from a BS to the rail & $d_0$ & 20 m  \\
Angle of the half-power beamwidth & $\theta_{-3dB}$ & $30^{\circ}$ \\
Shadowing margin &  $W$  & 10 dB \\
Path loss exponent & $n$ & 2 \\
Carrier wavelength & $\lambda$ & 5 mm \\
System bandwidth&  $B$ & 2.16 GHz \\
Noise figure of the receiver chain  &  $NF$   &  6 dB \\
Velocity &  $v$  & 300 km/h \\
Transmisson power  &   $P$ &    40 dBm\ and\ 50dBm \\
\hline
\end{tabular}
\label{table1}
\end{center}
\end{table}

We use the network geometry as shown in Fig. \ref{fig2}, and the simulation parameters are summarized in Table \ref{table1}. In this example, it is assumed that the position and velocity measurement is done at the edge of each BS's coverage, the power distribution is symmetrical.

\subsection{Comparison With Other Schemes}\label{S6-1}

In order to analyze the energy consumption of the scheme and further show the optimization results, a new scheme is introduced here, that is, combining the two power allocation methods, and four schemes are summarized as follows:

\textbf{Maintain constant transmission power (MCTP)}: Maintain constant transmission power $P$ throughout the process.

\textbf{Optimization transmission power allocation (OTPA)}: Allocate the transmission power rationally using power minimization algorithm.

\textbf{Mixed transmission power allocation (MTPA)}: Maintain constant transmit power $P$ within the range $[0,{d_l}/4]$, and then use the power minimization algorithm for power allocation within the range $[{d_l}/4,{d_l}/2]$.

\textbf{When $N$ tends to infinity (OTPA($N\to\infty$))}: Allocate the transmission power rationally by power minimization algorithm and then find the limit when $N$ tends to infinity by formula (\ref{eq40}).

MTPA is a hybrid optimization scheme and OTPA($N\to\infty$) is the limit value of OTPA. The energy consumption of four schemes is simulated as follows.

\begin{figure}[t]
\begin{center}
\includegraphics*[width=0.9\columnwidth,height=2.3in]{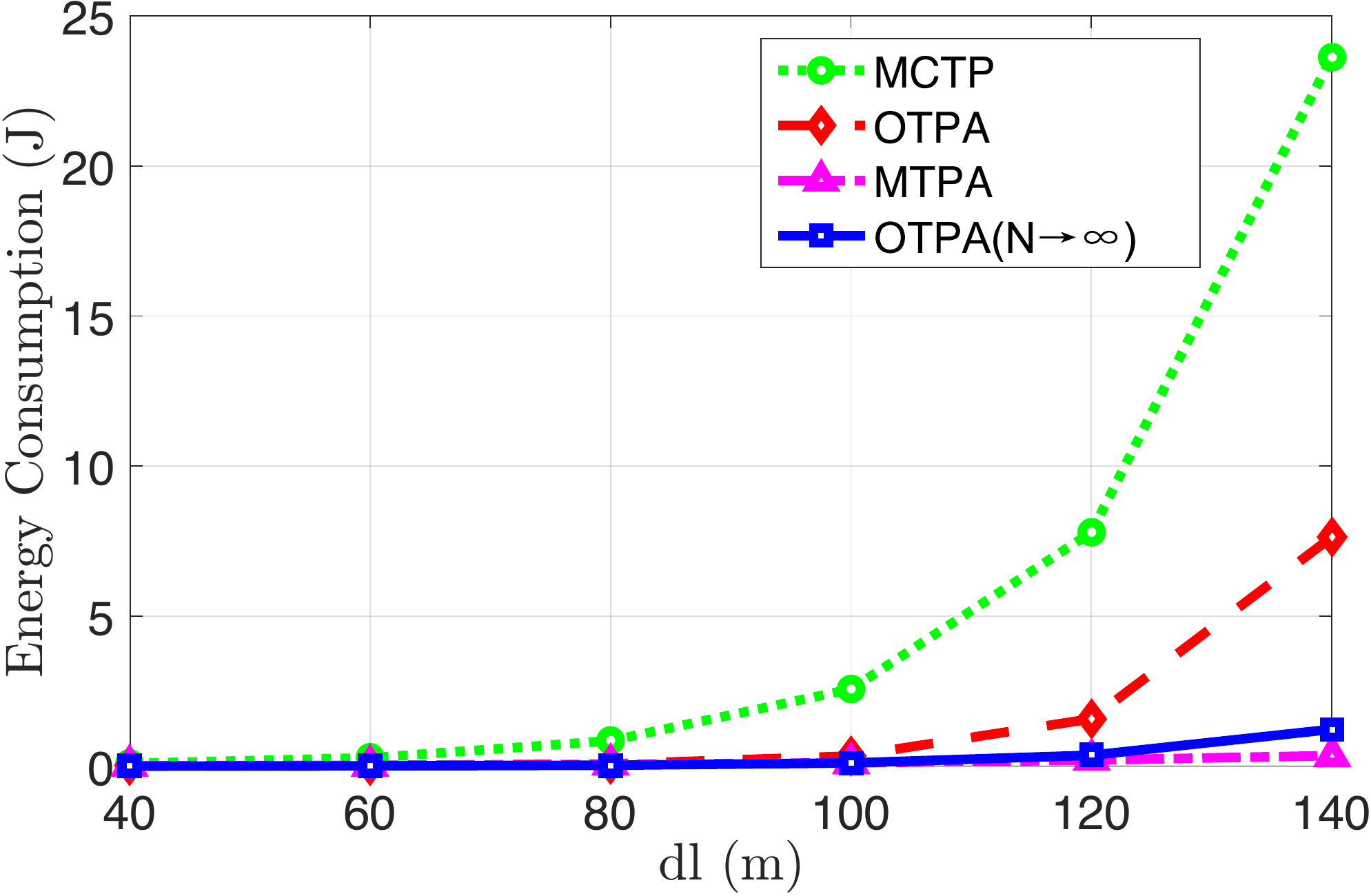}
\end{center}
\caption{Energy consumption comparison under different $d_l$\ $(P=40dBm)$.}
\label{fig5}
\end{figure}

\begin{figure}[t]
\begin{center}
\includegraphics*[width=0.9\columnwidth,height=2.3in]{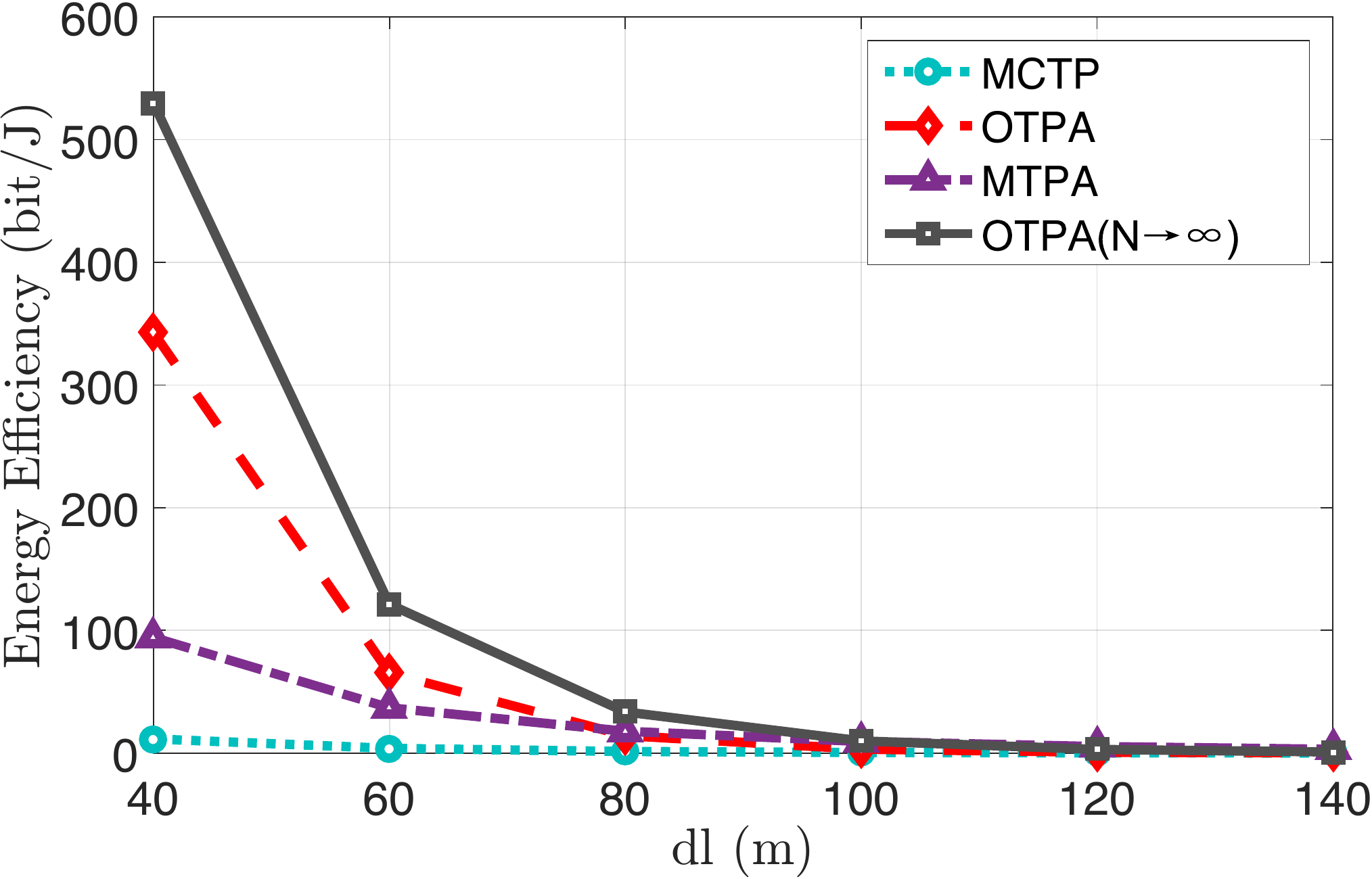}
\end{center}
\caption{Energy efficiency comparison under different $d_l$\ $(P=40dBm)$.}
\label{fig6}
\end{figure}

\begin{figure}[htp]
\begin{center}
\includegraphics*[width=0.9\columnwidth,height=2.3in]{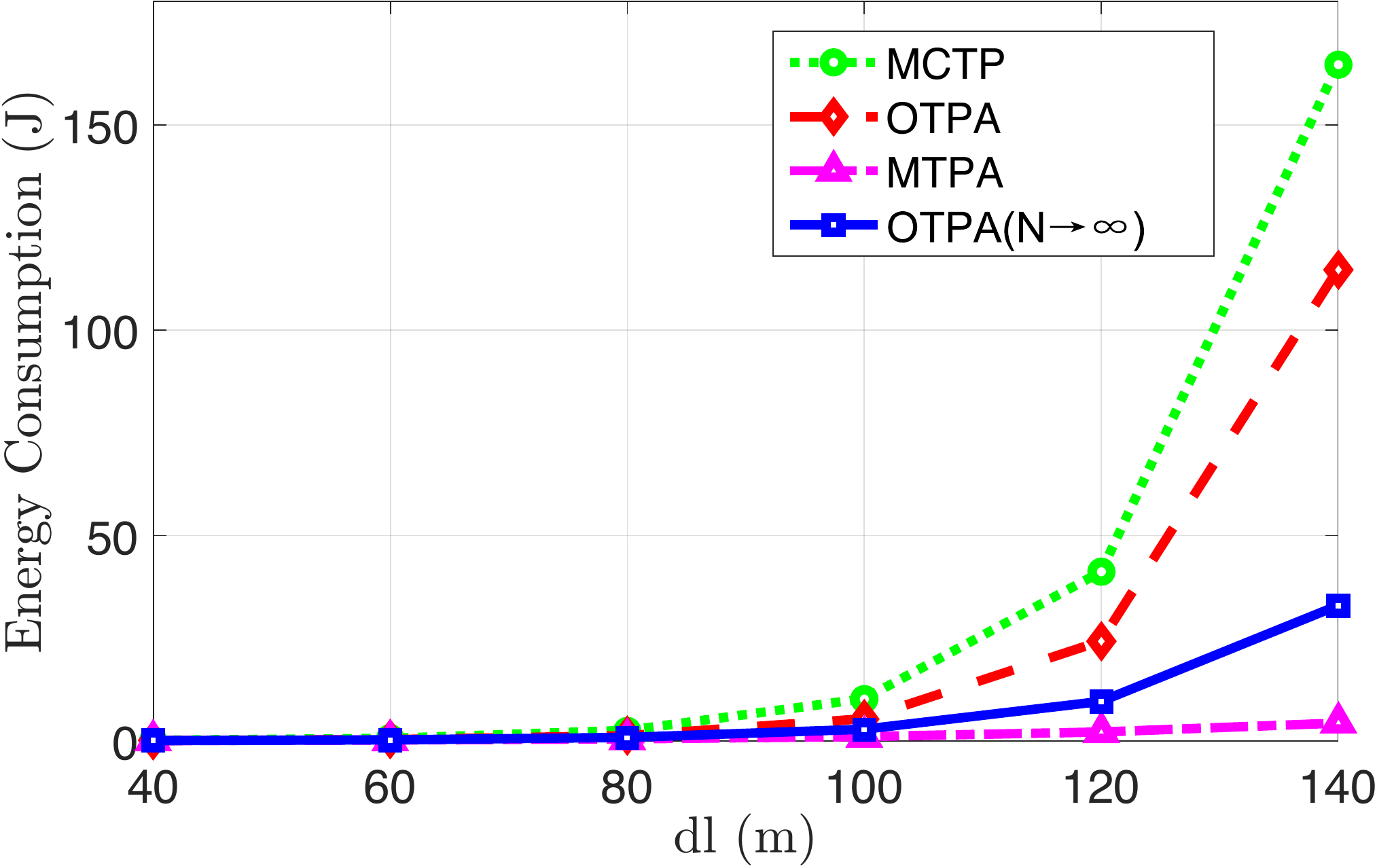}
\end{center}
\caption{Energy consumption comparison under different $d_l$\ $(P=50dBm)$.}
\label{fig7}
\end{figure}

\begin{figure}[htp]
\begin{center}
\includegraphics*[width=0.9\columnwidth,height=2.3in]{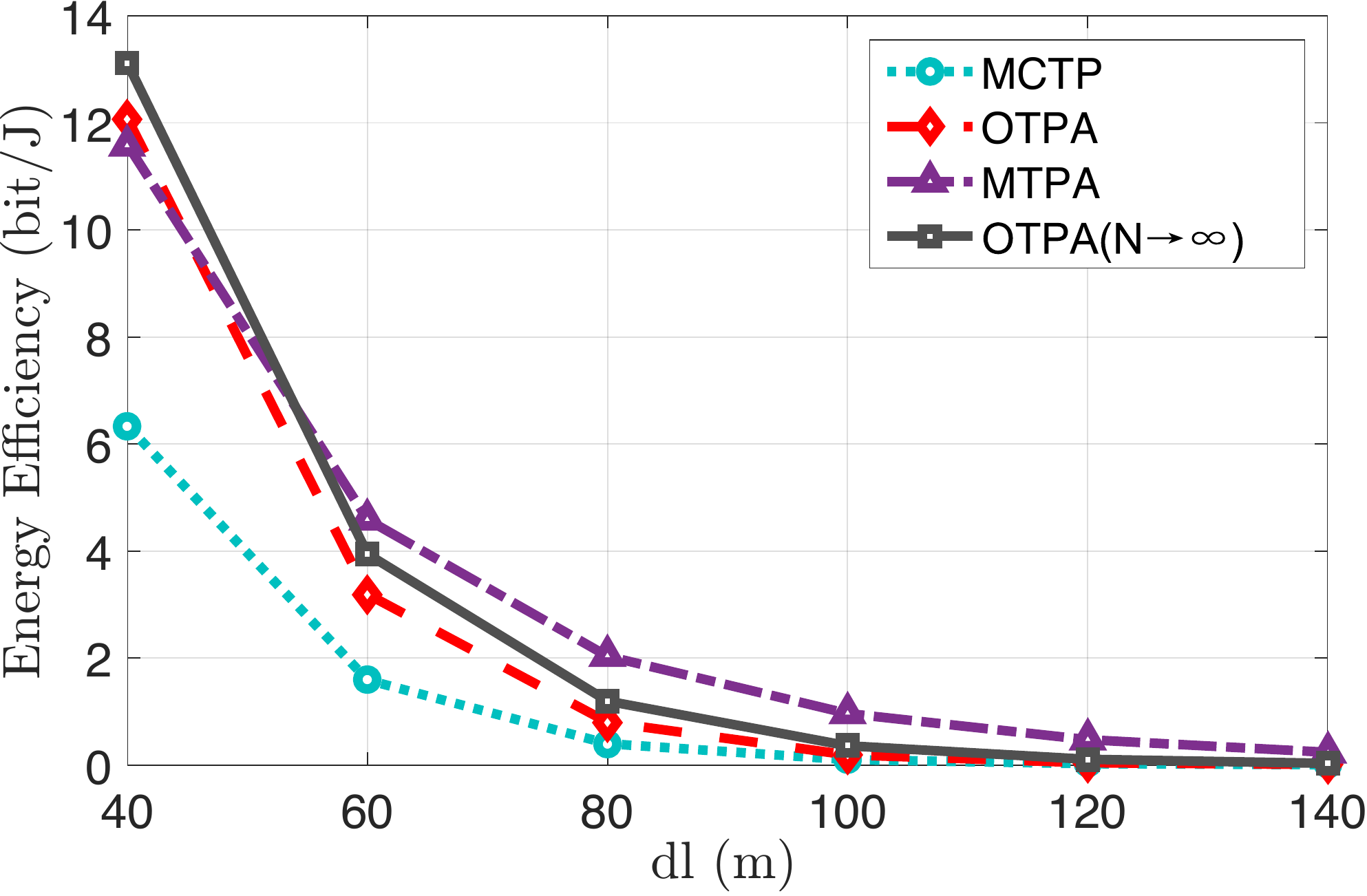}
\end{center}
\caption{Energy efficiency comparison under different $d_l$\ $(P=50dBm)$.}
\label{fig8}
\end{figure}

\begin{figure}[htp]
\begin{center}
\includegraphics*[width=0.9\columnwidth,height=2.3in]{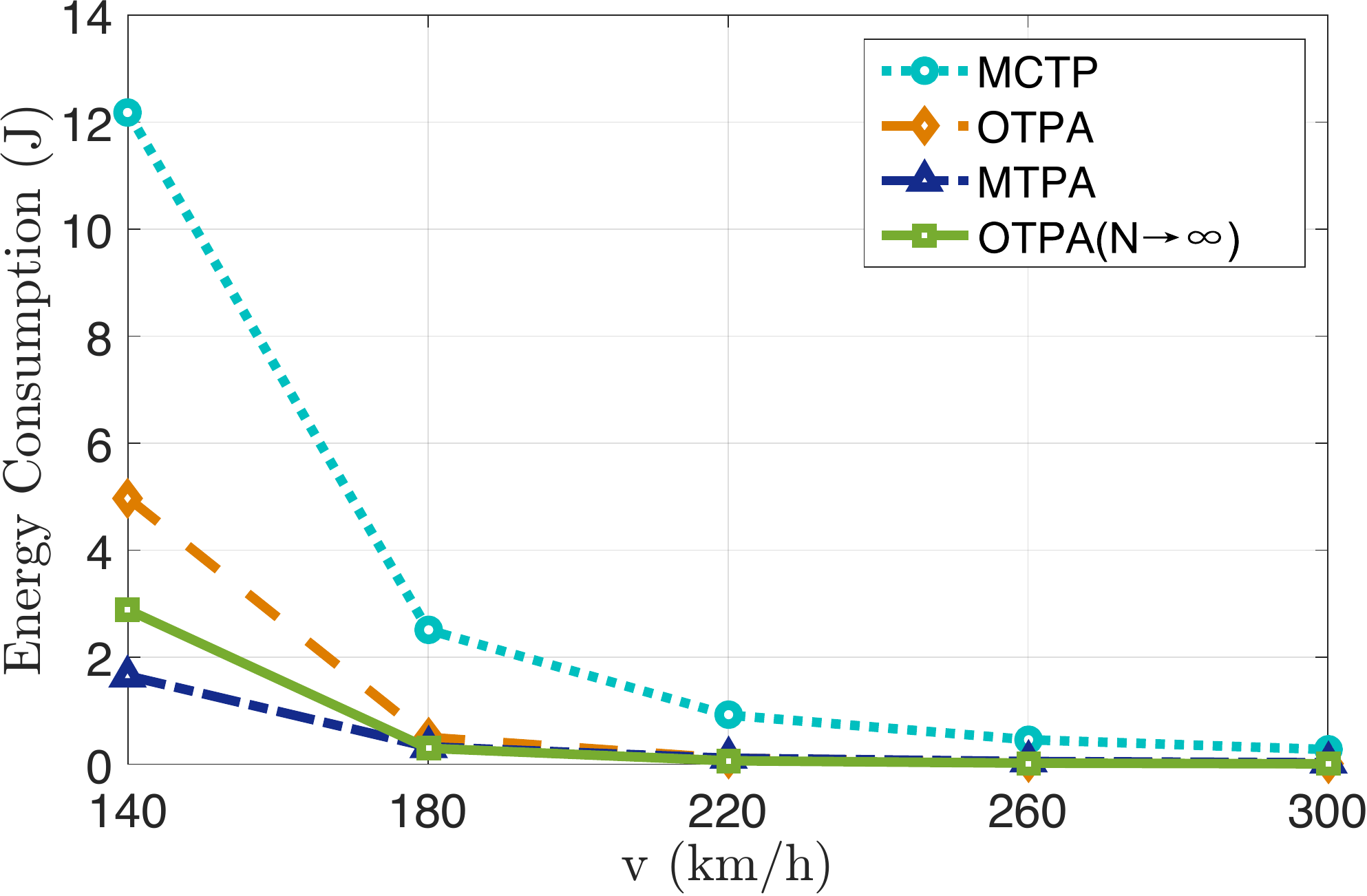}
\end{center}
\caption{Energy consumption comparison under different $v$\ $(P=40dBm)$.}
\label{fig15}
\end{figure}

\begin{figure}[htp]
\begin{center}
\includegraphics*[width=0.9\columnwidth,height=2.3in]{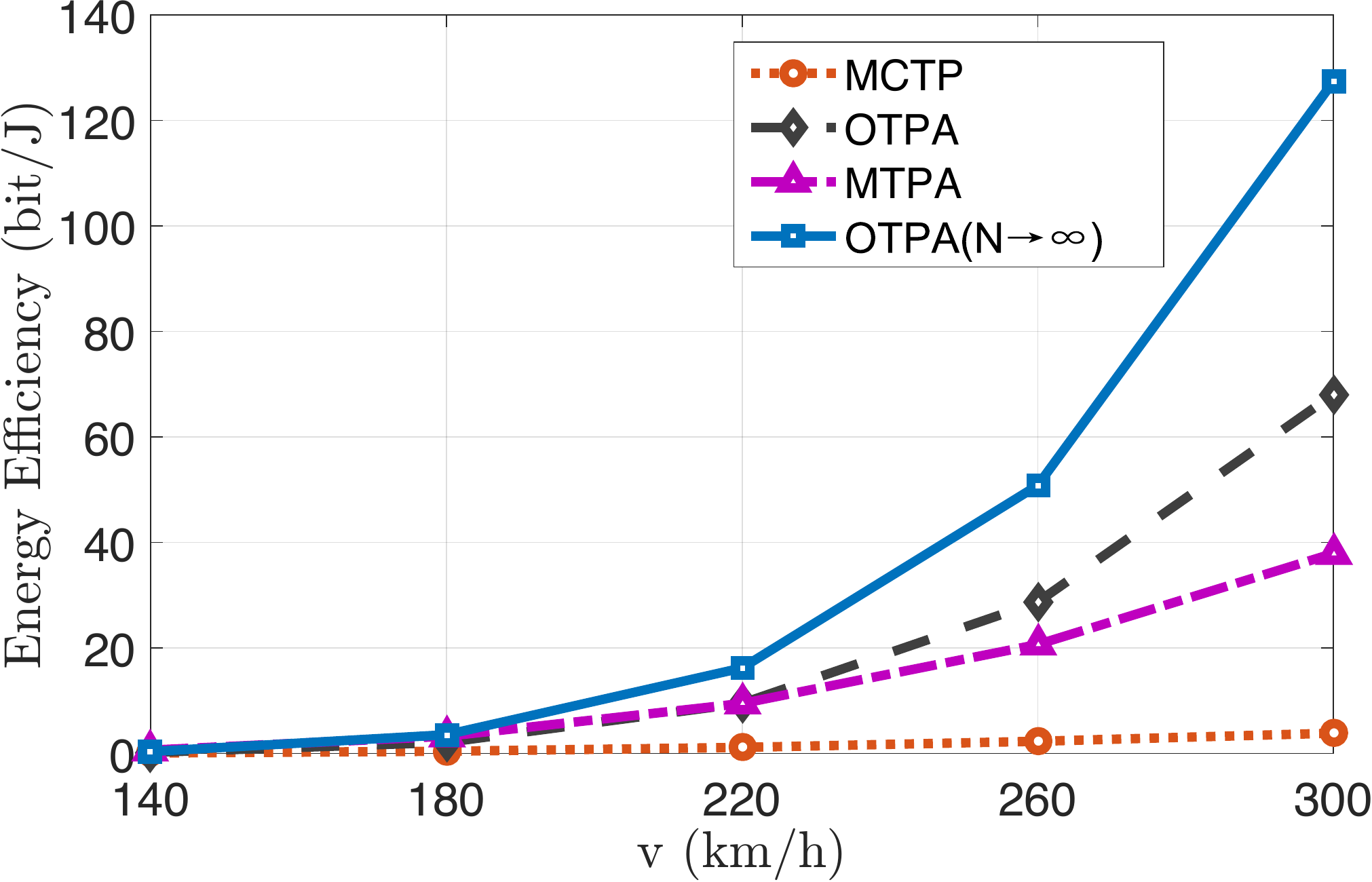}
\end{center}
\caption{Energy efficiency comparison under different $v$\ $(P=40dBm)$.}
\label{fig16}
\end{figure}

\begin{figure}[htp]
\begin{center}
\includegraphics*[width=0.9\columnwidth,height=2.3in]{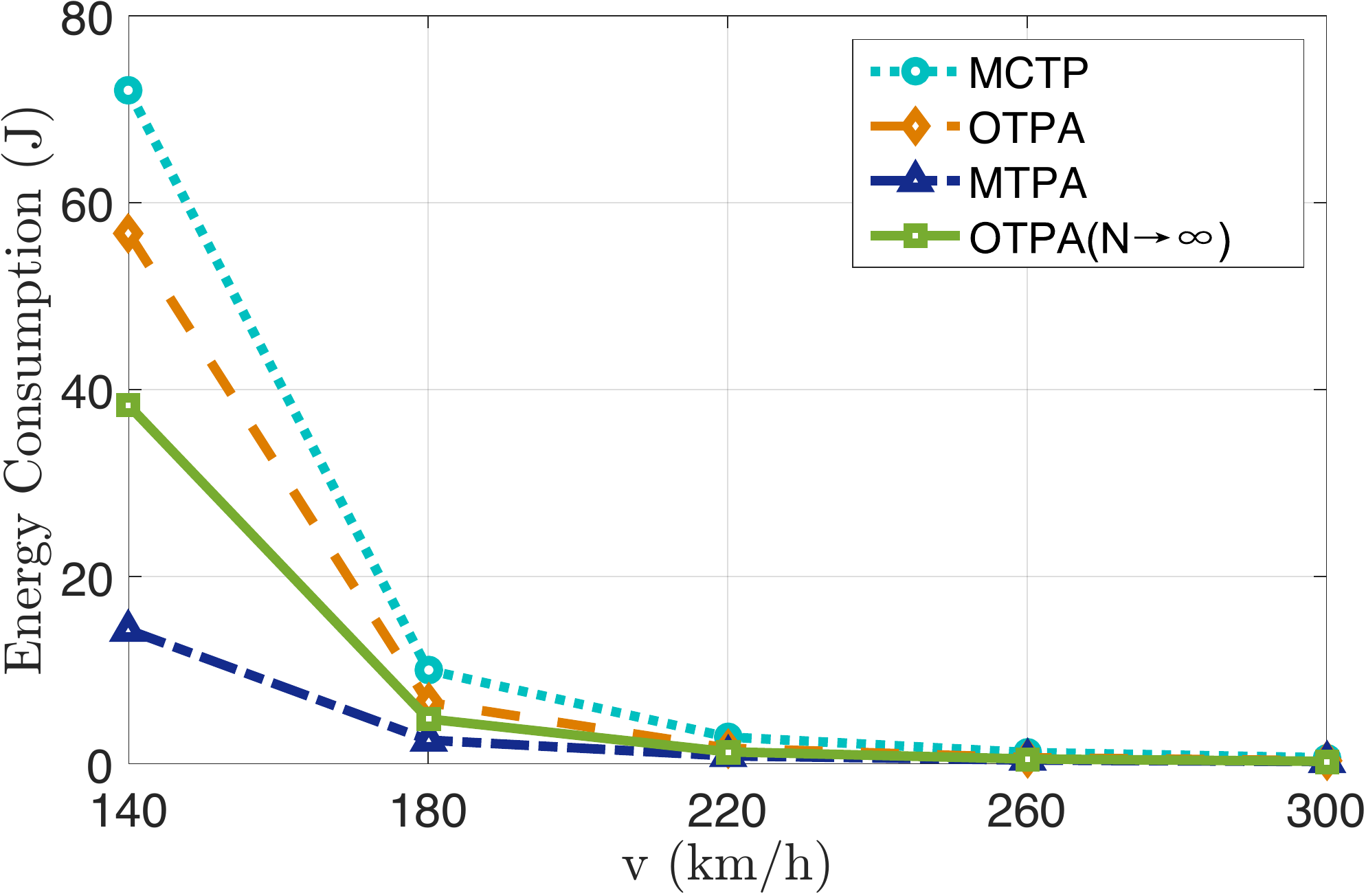}
\end{center}
\caption{Energy consumption comparison under different $v$\ $(P=50dBm)$.}
\label{fig17}
\end{figure}

\begin{figure}[htp]
\begin{center}
\includegraphics*[width=0.9\columnwidth,height=2.3in]{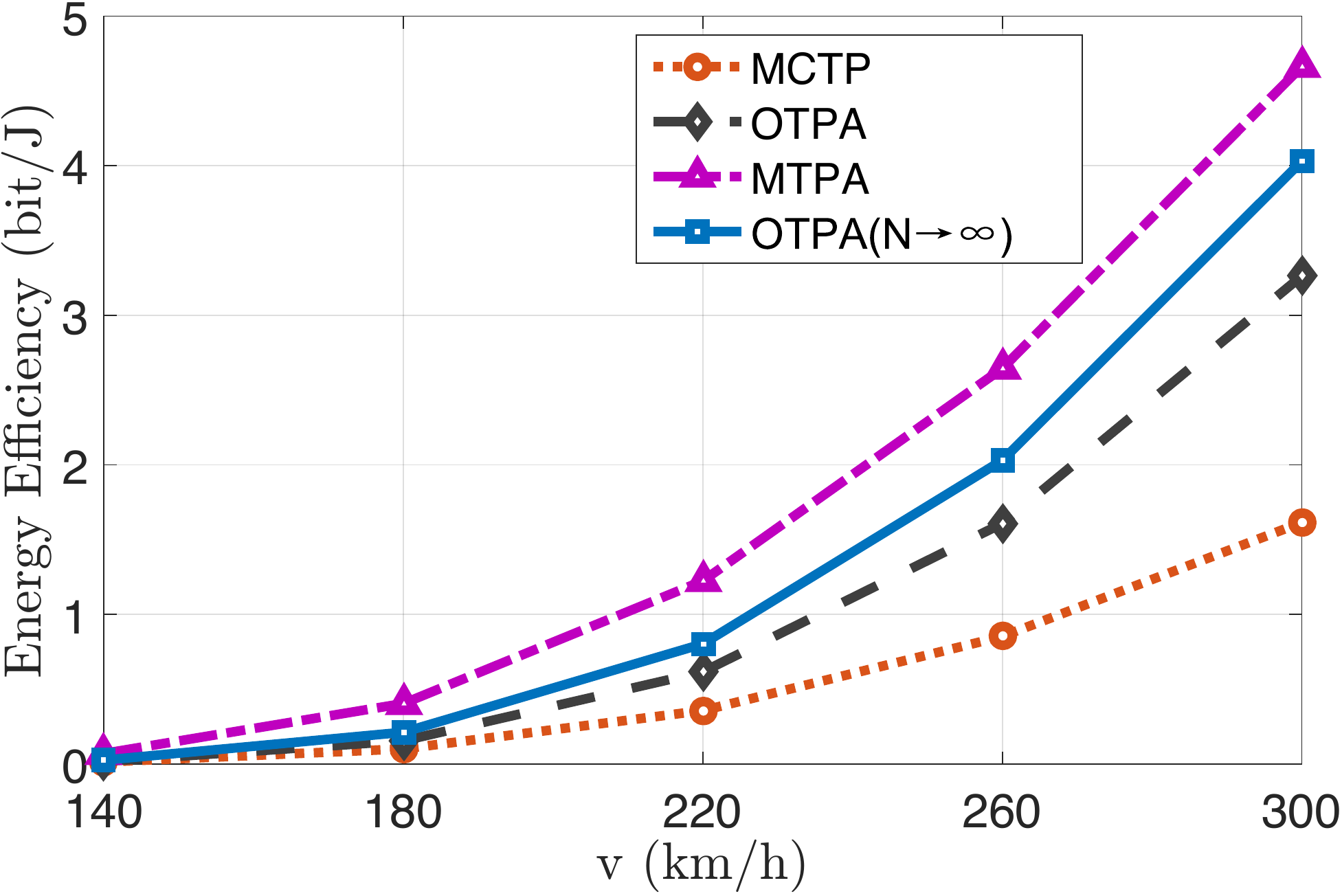}
\end{center}
\caption{Energy efficiency comparison under different $v$\ $(P=50dBm)$.}
\label{fig18}
\end{figure}

In Fig. \ref{fig5}, we plot the energy consumption comparison of four schemes under different inter-BS distances $d_l$. From the results, we can observe that energy consumption increases with the distance and the latter three optimization schemes have played a role in reducing energy consumption especially when $d_l>100m$. The energy consumption of MCTP is higher when distance is far and the growth rate will be larger when distance is farther, from this, we can also discover the importance of reducing energy consumtion. To be specific, when $d_l=140m$, OTPA saves about $67.7\%$ energy compared with MCTP. Compare OTPA and MTPA, the optimization effect of MTPA is better and can achieve lower energy consumption, but OTPA is also a good choice to some extent. Furthermore, we can see that the energy consumption of OTPA($N\to\infty$) is lower than OTPA, the result is as we calculated.

In Fig. \ref{fig6}, we plot the energy efficiency comparison of four schemes under different inter-BS distances $d_l$. Again, both OTPA and MTPA can achieve higher energy efficiency. When $d_l$ is small, the energy efficiency of OTPA is higher and the optimization effect is better compared with MTPA. In this situation, OTPA can be adopted as a green scheme. And with $d_l$ increases, the optimization effect of MTPA is gradually getting better. When $d_l>100m$, the energy efficiency of four schemes is approaching the same. Besides, the energy efficiency of OTPA($N\to\infty$) is the highest which verifies our calculation result.

We also plot the energy consumption comparison of four schemes when $P=50\ dBm$ in Fig. \ref{fig7}. Similarly, we can observe that all three optimization schemes reduce the energy consumption greatly and improve the energy efficiency of train-ground communication. The inter-BS distance is larger, the optimization effect is better. The energy efficiency comparison results are shown in Fig. \ref{fig8}.

In Fig. \ref{fig15} and Fig. \ref{fig17}, we plot the energy consumption comparison of four schemes under different velocity $v$. From the results, we can observe that our schemes can still achieve good performance in different speed situations, and energy consumption decreases with the increase of velocity. When $v=300km/h$, the value of latter three schemes are very small compared with MCTP, and the value gradually approaches zero because the distance is only $30m(d_l/2)$ in calculation which results in a small value. Compare Fig. \ref{fig15} and Fig. \ref{fig17}, we can see that when $P$ increases, energy consumption increases for a certain value of $v$. The energy efficiency comparison results are shown in Fig. \ref{fig16} and Fig. \ref{fig18}. It can be seen that when $P=40dBm$, the energy efficiency of OTPA is higher than MTPA, but the situation is reversed when $P=50dBm$.

In summary, our schemes consume lower energy and can achieve higher energy efficiency under different system parameters.

\subsection{Velocity Estimation Error}\label{S6-2}

Since the velocity estimation error has an important impact on the performance of beam switching, we now investigate it in the same method. Here, we assume the velocity estimation error is Gaussian, \emph{i.e.} $\hat{v}=v+v_e$, where $v$ is the true velocity and $v_e\sim\mathcal  N(0,\sigma_v^2)$.

\begin{figure}[htp]
\begin{center}
\includegraphics*[width=0.9\columnwidth,height=2.3in]{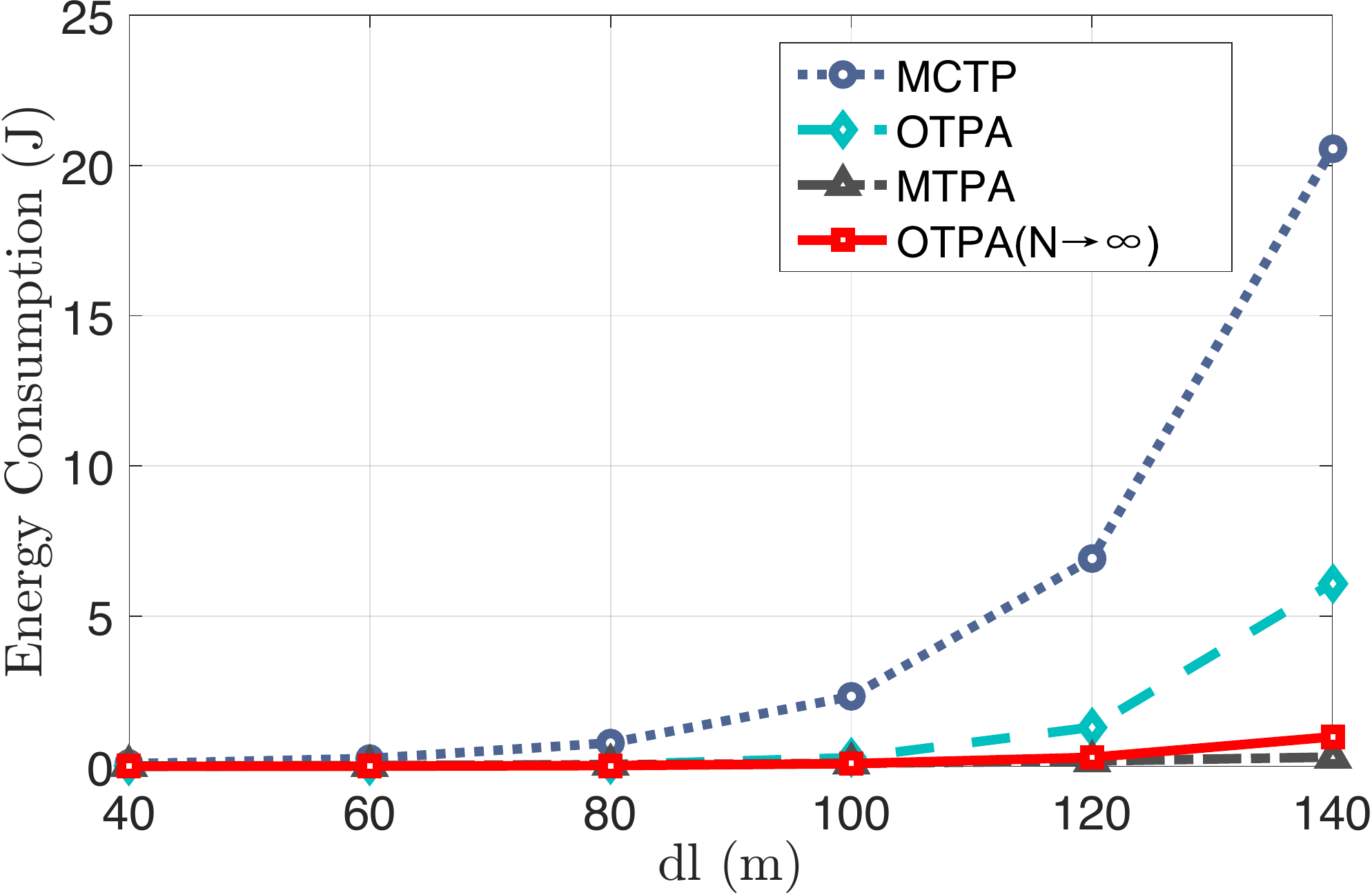}
\end{center}
\caption{Energy consumption comparison under different $d_l$ when $\sigma_v^2=(0.01v)^2$.}
\label{fig9}
\end{figure}

\begin{figure}[htp]
\begin{center}
\includegraphics*[width=0.9\columnwidth,height=2.3in]{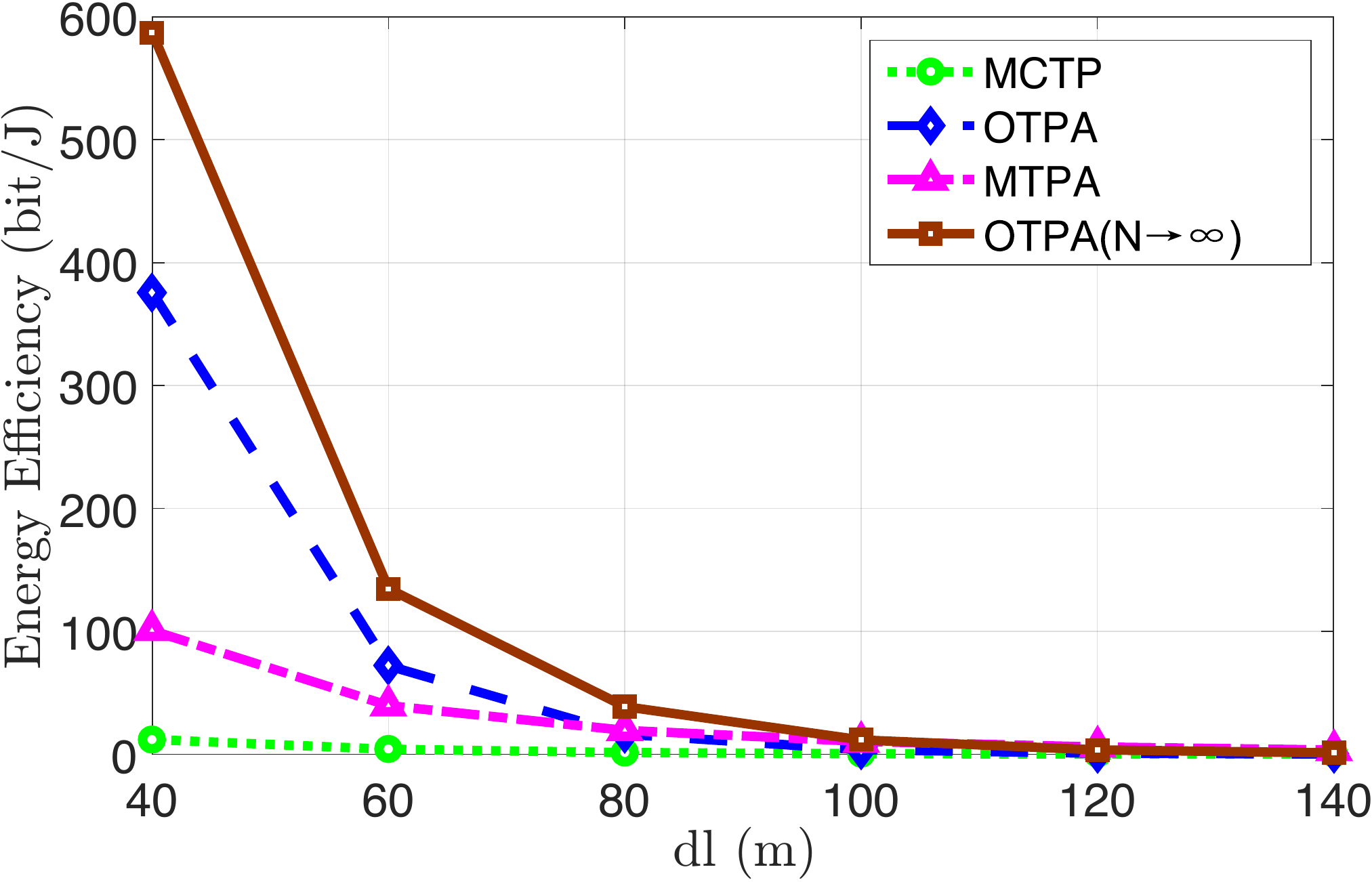}
\end{center}
\caption{Energy efficiency comparison under different $d_l$ when $\sigma_v^2=(0.01v)^2$.}
\label{fig10}
\end{figure}

\begin{figure}[htp]
\begin{center}
\includegraphics*[width=0.9\columnwidth,height=2.3in]{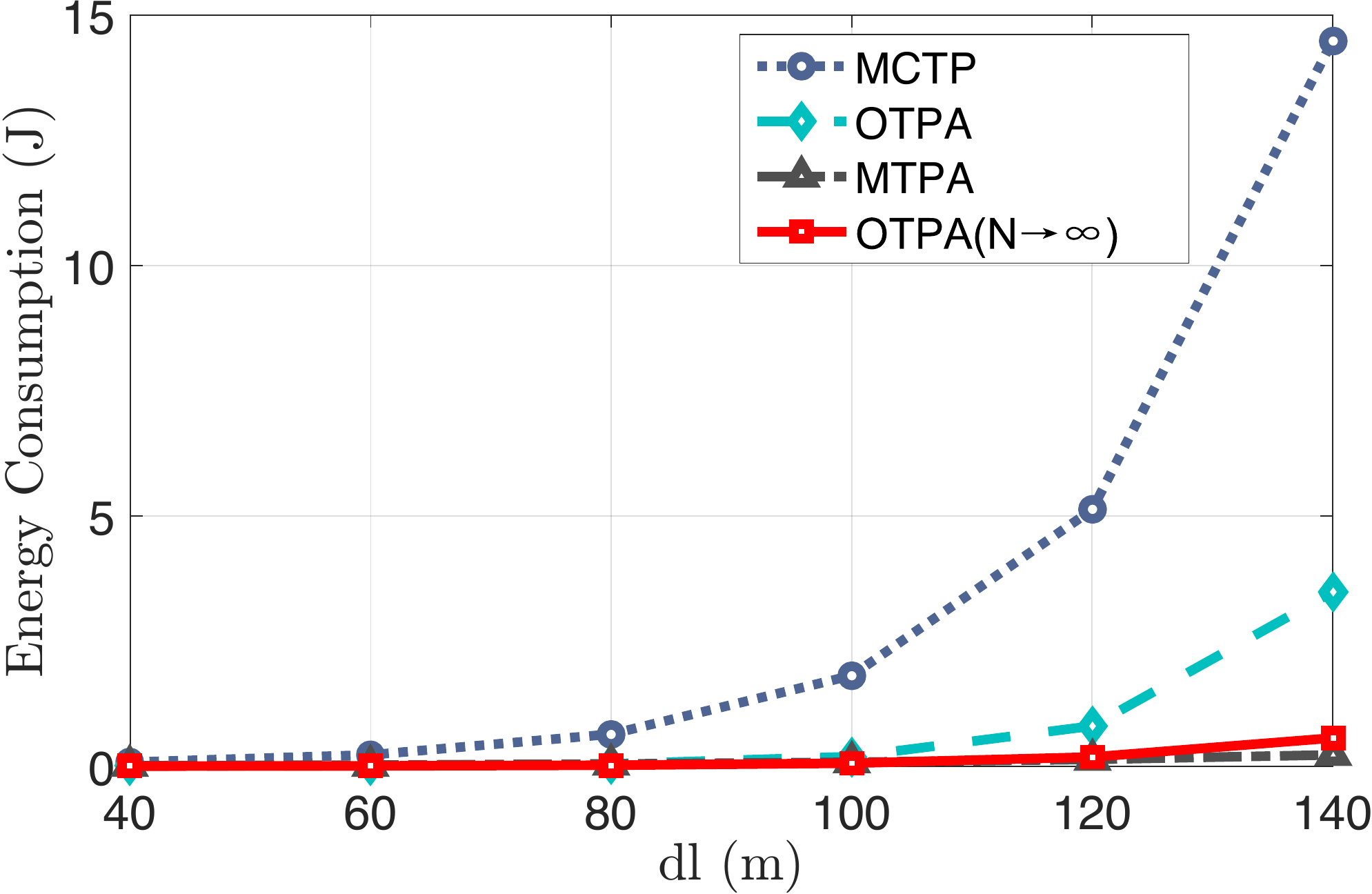}
\end{center}
\caption{Energy consumption comparison under different $d_l$ when $\sigma_v^2=(0.1v)^2$.}
\label{fig11}
\end{figure}

\begin{figure}[htp]
\begin{center}
\includegraphics*[width=0.9\columnwidth,height=2.3in]{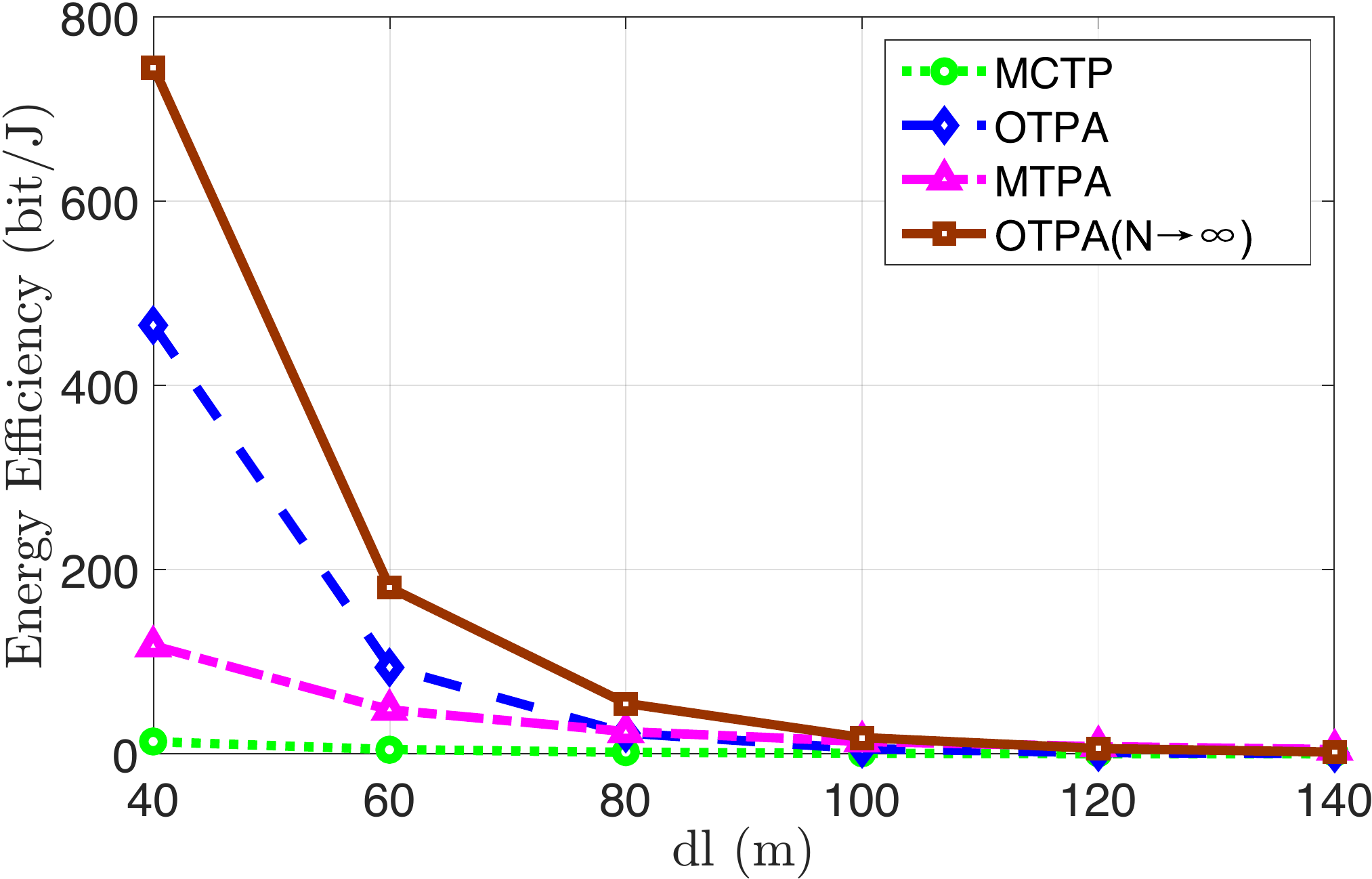}
\end{center}
\caption{Energy efficiency comparison under different $d_l$ when $\sigma_v^2=(0.1v)^2$.}
\label{fig12}
\end{figure}

\begin{figure}[htp]
\begin{center}
\includegraphics*[width=0.9\columnwidth,height=2.3in]{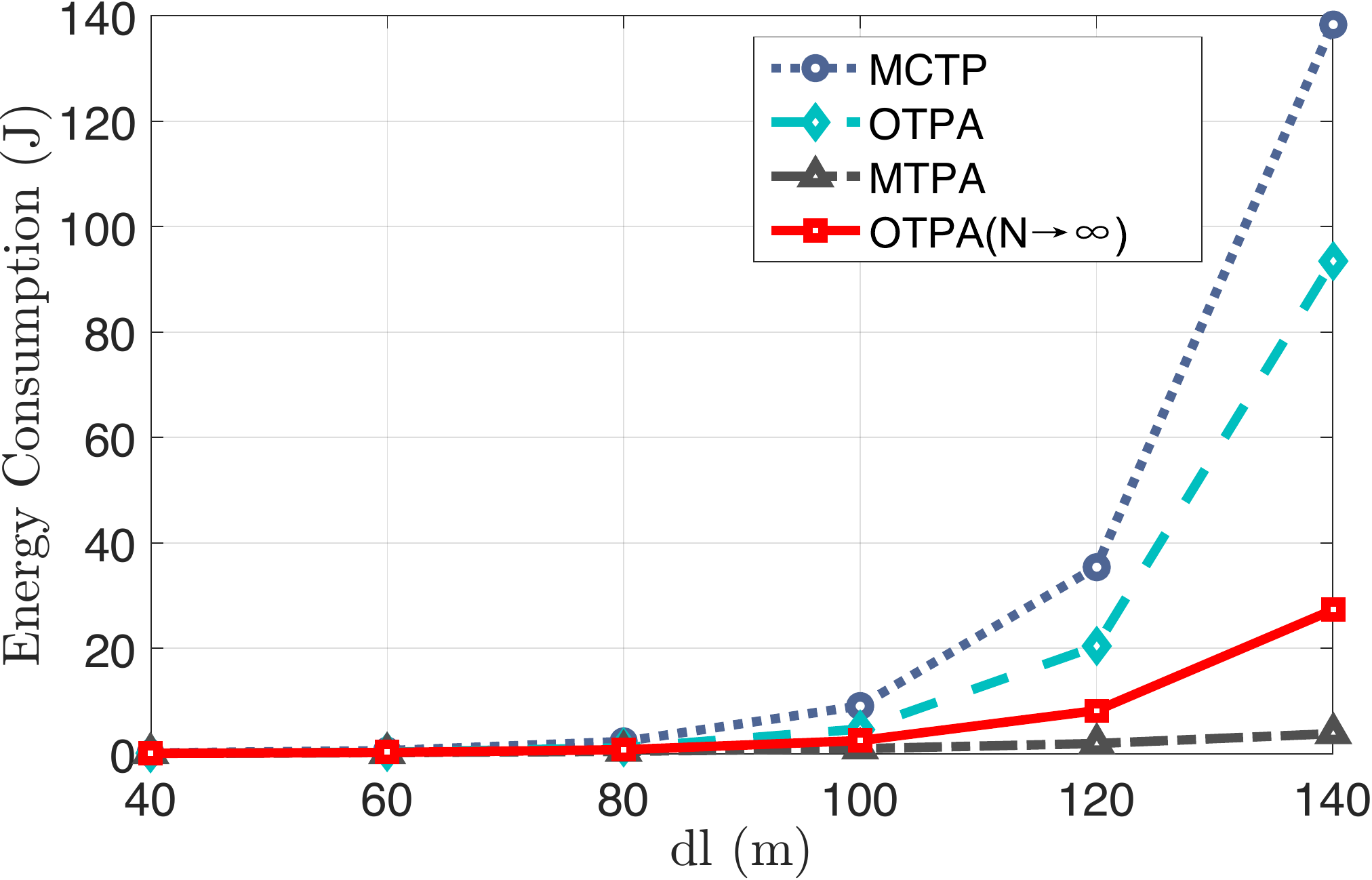}
\end{center}
\caption{Energy consumption comparison under different $d_l$ when $\sigma_v^2=(0.01v)^2$\quad $(P=50dBm)$.}
\label{fig13}
\end{figure}

\begin{figure}[htp]
\begin{center}
\includegraphics*[width=0.9\columnwidth,height=2.3in]{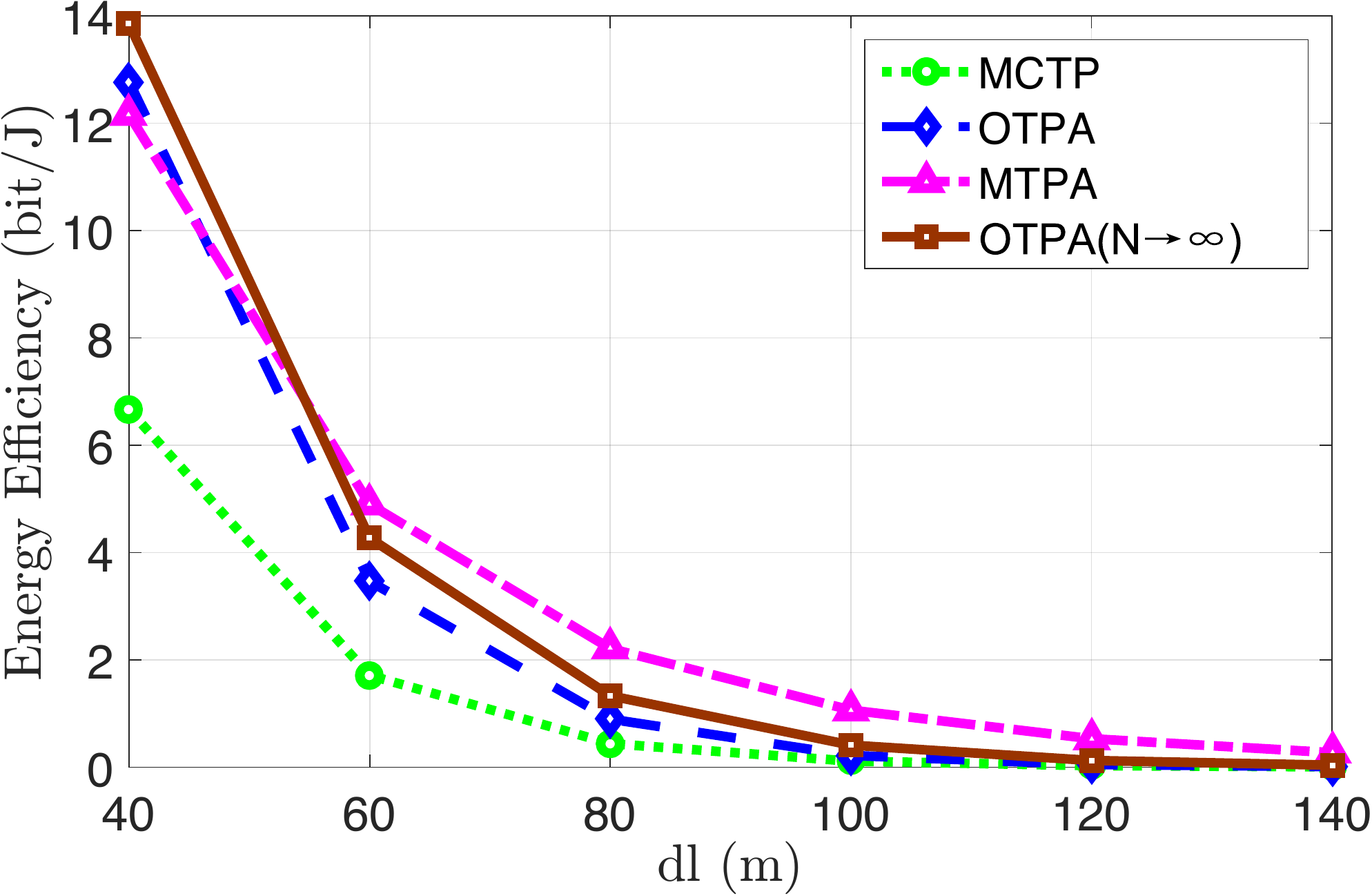}
\end{center}
\caption{Energy efficiency comparison under different $d_l$ when $\sigma_v^2=(0.01v)^2$\quad $(P=50dBm)$.}
\label{fig14}
\end{figure}

We plot the energy consumption comparison of four schemes under different $d_l$ when $\sigma_v^2=(0.01v)^2$ and $\sigma_v^2=(0.1v)^2$ in Fig. \ref{fig9} and Fig. \ref{fig11}. We can obtain that when take velocity estimation error into account, our schemes can also achieve lower energy consumption similar to the result in Fig. \ref{fig5}, where energy consumption increases with the increase of $d_l$. Through comparison, we can also observe that the energy consumption of four schemes in Fig. \ref{fig11} is lower than those in Fig. \ref{fig9}, it is because that the value of velocity $v$ in Fig. \ref{fig11} is bigger than that in Fig. \ref{fig9} as the result in  Fig. \ref{fig15}. Besides, it should be noted that the energy consumption of OTPA($N\to\infty$) is always lower than OTPA as we expected.

We also plot the energy efficiency comparison of four schemes under different $d_l$ in Fig. \ref{fig10} and Fig. \ref{fig12}. As we can  see, OTPA, MTPA and OTPA($N\to\infty$) can also perform well with velocity estimation error similar to the result in Fig. \ref{fig6}. Without loss of generality, we plot the energy consumption comparison and energy efficiency comparison of four schemes when $P=50dbm$ in Fig. \ref{fig13} and Fig. \ref{fig14}.

\section{Conclusion}\label{S7}

MmWave communication has the potential to solve the problem of train-ground communication for HSTs. Considering a beam switching method based on the position information of HSTs, this paper proposes an energy efficiency power control scheme where the transmission power is allocated rationally through the power minimization algorithm. We also develop a hybrid optimization scheme and the situation which the limit of the number of segments tends to infinity is considered. Extensive simulations have demonstrate that our schemes can achieve lower energy consumption and higher energy efficiency. In the future work, we will consider a specific channel model and evaluate the energy efficiency of the scheme where the Doppler effect has been eliminated.

\bibliographystyle{IEEEtran}

\vspace*{58mm}
\begin{IEEEbiography}[{\includegraphics[width=1in,height=1.25in,clip,keepaspectratio]{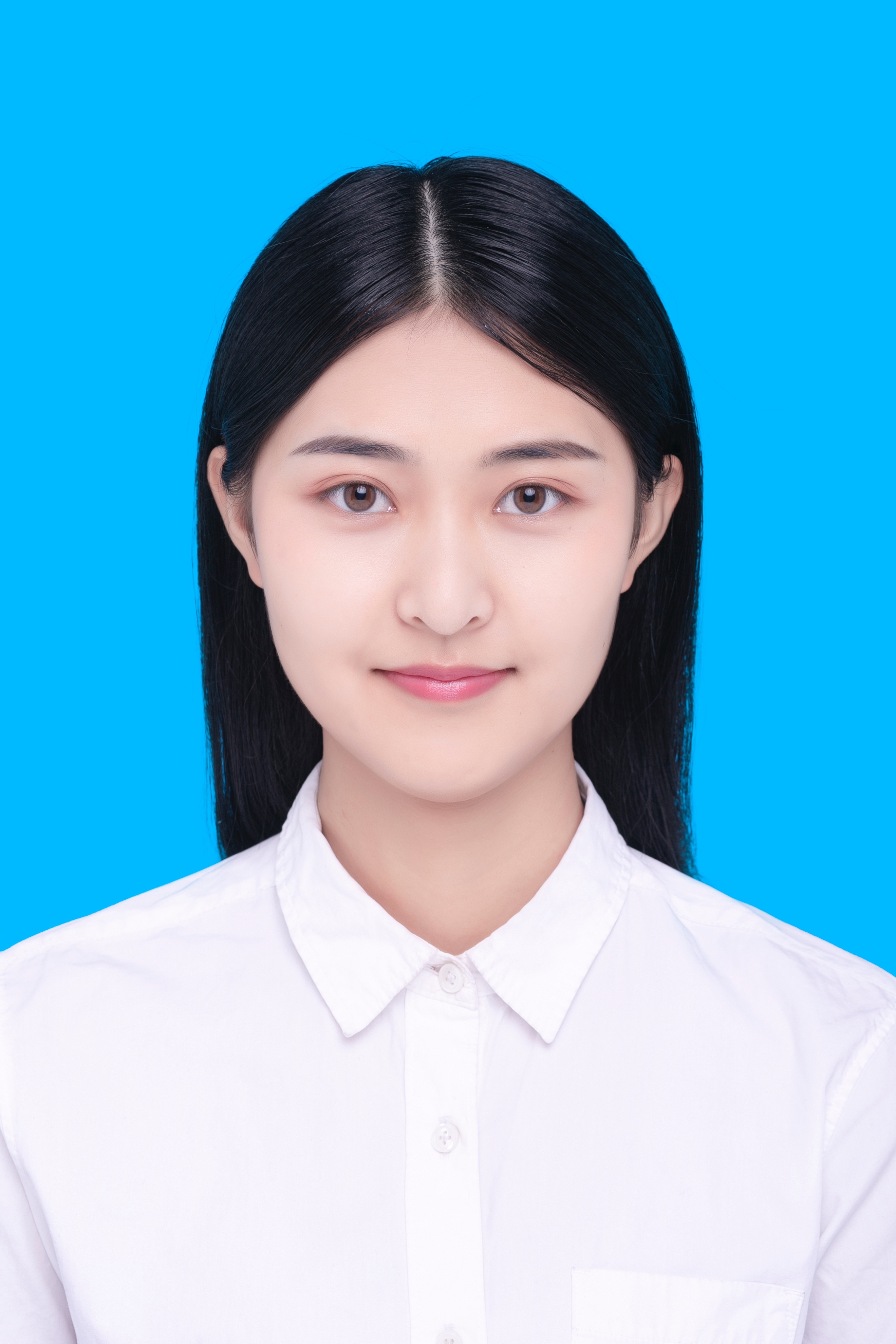}}]{Lei Wang}
was born in Shanxi, China, in 1997. She received the B.S. degree in mathematics and applied mathematics from Beijing Jiaotong University, Beijing, China, in 2019. She is currently pursuing the M.S. degree with the State Key Laboratory of Rail Traffic Control
and Safety, Beijing Jiaotong University, Beijing, China. Her research interest is millimeter-wave wireless communications.
\vspace*{32mm}
\end{IEEEbiography}

\begin{IEEEbiography}[{\includegraphics[width=1in,height=1.25in,clip,keepaspectratio]{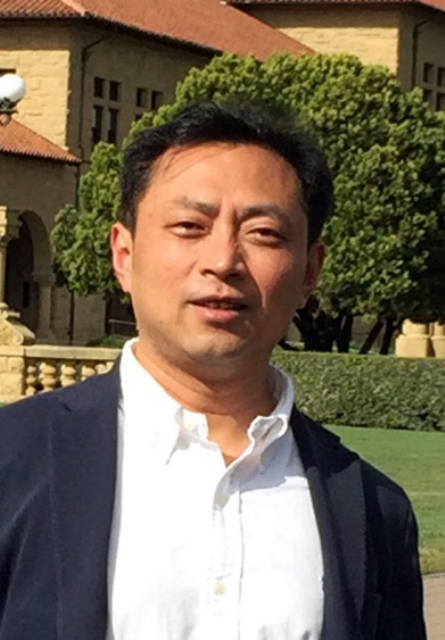}}]{Bo Ai}
received the M.S. and Ph.D. degrees from Xidian University, China. He studies as a Post-Doctoral Student at Tsinghua University. He was a Visiting Professor with the Electrical Engineering Department, Stanford University, in 2015. He is currently with Beijing Jiaotong University as a Full Professor and a Ph.D. Candidate Advisor. He is the Deputy Director of the State Key Lab of Rail Traffic Control and Safety and the Deputy Director of the International Joint Research Center. He is one of the main people responsible for the Beijing Urban Rail Operation Control System, International Science and Technology Cooperation Base. He is also a Member, of the Innovative Engineering Based jointly granted by the Chinese Ministry of Education and the State Administration of Foreign Experts Affairs. He was honored with the Excellent Postdoctoral Research Fellow by Tsinghua University in 2007.

He has authored/co-authored eight books and published over 300 academic research papers in his research area. He holds 26 invention patents. He has been the research team leader for 26 national projects. His interests include the research and applications of channel measurement and channel modeling, dedicated mobile communications for rail traffic systems. He has been notified by the Council of Canadian Academies that, based on Scopus database, he has been listed as one of the Top 1\% authors in his field all over the world. He has also been feature interviewed by the IET Electronics Letters. He has received some important scientific research prizes.

Dr. Ai is a fellow of the Institution of Engineering and Technology. He is an Editorial Committee Member of the Wireless Personal Communications journal. He has received many awards, such as the Outstanding Youth Foundation from the National Natural Science Foundation of China, the Qiushi Outstanding Youth Award by the Hong Kong Qiushi Foundation, the New Century Talents by the Chinese Ministry of Education, the Zhan Tianyou Railway Science and Technology Award by the Chinese Ministry of Railways, and the Science and Technology New Star by the Beijing Municipal Science and Technology Commission. He was a co-chair or a session/track chair for many international conferences. He is an IEEE VTS Beijing Chapter Vice Chair and an IEEE BTS Xi'an Chapter Chair. He is the IEEE VTS Distinguished Lecturer. He is an Editor of the IEEE TRANSACTIONS ON CONSUMER ELECTRONICS. He is the Lead Guest Editor of Special Issues of the IEEE TRANSACTIONS ON VEHICULAR TECHNOLOGY, the IEEE ANTENNAS AND WIRELESS PROPAGATION LETTERS, and the International Journal of Antennas and Propagation.
\end{IEEEbiography}

\begin{IEEEbiography}[{\includegraphics[width=1in,height=1.25in,clip,keepaspectratio]{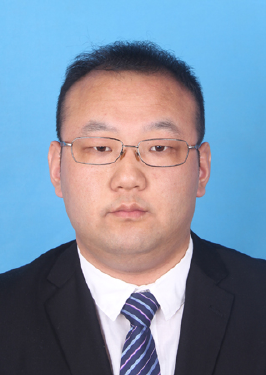}}]{Yong Niu}
(M'17) received the B.E. degree in Electrical Engineering from Beijing Jiaotong University, China, in 2011, and the Ph.D. degree in Electronic Engineering from Tsinghua University, Beijing, China, in 2016.

From 2014 to 2015, he was a Visiting Scholar with the University of Florida, Gainesville, FL, USA. He is currently an Associate Professor with the State Key Laboratory of Rail Traffic Control and Safety, Beijing Jiaotong University. His research interests are in the areas of networking and communications, including millimeter wave communications, device-to-device communication, medium access control, and software-defined networks. He received the Ph.D. National Scholarship of China in 2015, the Outstanding Ph.D. Graduates and Outstanding Doctoral Thesis of Tsinghua University in 2016, the Outstanding Ph.D. Graduates of Beijing in 2016, and the Outstanding Doctorate Dissertation Award from the Chinese Institute of Electronics in 2017. He has served as Technical Program Committee member for IWCMC 2017, VTC2018-Spring, IWCMC 2018, INFOCOM 2018, and ICC 2018. He was the Session Chair for IWCMC 2017. He was the recipient of the 2018 International Union of Radio Science Young Scientist Award.
\end{IEEEbiography}

\begin{IEEEbiography}[{\includegraphics[width=1in,height=1.25in,clip,keepaspectratio]{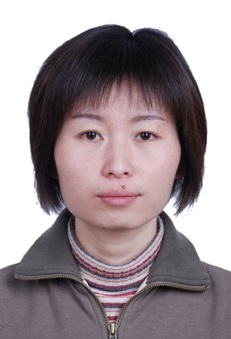}}]{Xia Chen}
(xchen@bjtu.edu.cn) received her B.Eng. and Ph.D. degrees from Beijing Jiaotong University in 1997 and 2003, respectively. She has been with Beijing Jiaotong University since 2003, where she is currently an associate professor with the School of Electronic and Information Engineering. Her research interests include mobile channel modeling, multicarrier transmission, and radio resource management of wireless networks. She is a member of the Chinese Institute of Electronics.
\end{IEEEbiography}

\begin{IEEEbiography}[{\includegraphics[width=1in,height=1.25in,clip,keepaspectratio]{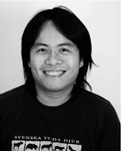}}]{Pan Hui}
(M'04SM'14F'18) received the B.Eng. and M.Phil. degrees from the Department of Electrical and Electronic Engineering, The University of Hong Kong, and the Ph.D. degree from the Computer Laboratory, University of Cambridge. He was a Senior Research Scientist and a Distinguished Scientist for Telekom Innovation Laboratories, Germany, from 2008 to 2015. He has been an Adjunct Professor of social computing and networking with Aalto University, Finland, since 2012, and a Faculty Member with the Department of Computer Science and Engineering, The Hong Kong University of Science and Technology, since 2013. Since 2017, he has been the Nokia Chair of data science and a Full Professor of computer science with the University of Helsinki. During his Ph.D. period, he was with Intel Research Cambridge and Thomson Research Paris. He has published over 250 research papers with over 16,000 citations and has around 30 granted/led European patents. He has founded and chaired several IEEE/ACM conferences/workshops. He has been serving on the Organizing and Technical Program Committee of numerous international conferences, including the ACM SIGCOMM, IEEE Infocom, ICNP, SECON, MobiSys, AAAI, IJCAI, ICWSM, and WWW. He is an Associate Editor of the IEEE TRANSACTIONS ON MOBILE COMPUTING and was an Associate Editor of the IEEE TRANSACTIONS ON CLOUD COMPUTING. He is an ACM Distinguished Scientist.
\end{IEEEbiography}

\end{document}